    \documentclass{aa}  

\usepackage{txfonts}
\usepackage{graphicx,lipsum}
\usepackage[percent]{overpic}
\usepackage{amssymb,amsmath}
\usepackage[version=4]{mhchem}
\usepackage{newtxtext}
\usepackage[varvw]{newtxmath}
\usepackage{xcolor}
\usepackage{caption,subcaption}

\newcommand{\ignore}[1]{}

\def\aap {{A\&A}}

\def\apjl {{ApJL}}

\def\apjs {{ApJS}}

\begin{document}

   \title{ Gas dynamics around a Jupiter mass planet: II. Chemical evolution of circumplanetary material }
   \titlerunning{ Chemical evolution of circumplanetary material }


   \author{Alex J. Cridland\inst{1,2,3}\thanks{A.Cridland@lmu.de}, Elena Lega\inst{1}, Myriam Benisty\inst{1,2,4} 
          }
    \authorrunning{ Cridland, Lega, \& Benisty }

   \institute{
   $^1$Université Côte d'Azur, Observatoire de la Côte d'Azur, CNRS, Laboratoire Lagrange, 96 Bd. de l'Observatoire, 06300 Nice, France\\
   $^2$Univ. Grenoble Alpes, CNRS, IPAG, 414 Rue de la Piscine, 38000 Grenoble, France\\
   $^3$Universit\"ats-Sternwarte M\"unchen, Ludwig-Maximilians-Universit\"at, Scheinerstr. 1, 81679 M\"unchen, Deutschland\\
   $^4$Max-Planck Institute for Astronomy (MPIA), Königstuhl 17, 69117 Heidelberg, Deutschland
             }

   \date{Received \today}


  \abstract
  {
  In an ongoing effort to understand planet formation the link between the chemistry of the protoplanetary disk and the properties of resulting planets have long been a subject of interest. These connections have generally been made between mature planets and young protoplanetary disks through the carbon-to-oxygen (C/O) ratio. In a rare number of systems, young protoplanets have been found within their natal protoplanetary disks. These systems offer a unique opportunity to directly study the delivery of gas from the protoplanetary disk to the planet. In this work we post-process 3D numerical simulations of an embedded Jupiter-massed planet in its protoplanetary disk to explore the chemical evolution of gas as it flows from the disk to the planet. The relevant dust to this chemical evolution is assumed to be small, co-moving grains with a reduced dust-to-gas ratio indicative of the upper atmosphere of a protoplanetary disk. We find that as the gas enters deep into the planet's gravitational well, it warms significantly (up to $\sim 800$ K), releasing all of the volatile content from the ice phase. This change in phase can influence our understanding of the delivery of volatile species to the atmospheres of giant planets. The primary carbon, oxygen, and sulfur carrying ices: CO$_2$, H$_2$O, and H$_2$S are released into the gas phase and along with the warm gas temperatures near the embedded planets lead to the production of unique species like CS, SO, and SO$_2$ compared to the protoplanetary disk. We compute the column densities of SO, SO$_2$, CS, and H$_2$CS 
  in our model and find that their values 
  are consistent with previous observational studies.
}


    \maketitle
%

\section{Introduction}\label{sec:intro}


The gravity sculpted gap in the protoplanetary disk (PPD) gas and dust opens in the final stages of a giant planet's growth. In this stage, the planet has become sufficiently massive that its gravitational influence exceeds the disk viscosity at the midplane, clearing away the co-orbiting material. Gas flows from the PPD towards the embedded protoplanet through the first and second Lagrange points, and from heights above the midplane \citep[the so-called meridional flows,][]{Morbidelli2014,Szulagyi2016}.

There is a general expectation that this late stage gas flow will collect into a circumplanetary disk (CPD) because of angular momentum conservation and analogies to star formation. Further evidence of the co-planar Galilean moons of Jupiter suggest that they formed in an analogous ways to the rest of the planets around our Sun \citep[i.e. in a disk-like structure around Jupiter,][]{CanupWard2006,CanupWard2009}. Observing these CPDs (or any circumplanetary material in general) has been primarily attempted through dust emission observation and remains a challenging endeavour due to the possible contamination from surrounding PPD material. As such there are only a few observational works which suggest the presence of a CPD, around the PDS70 protoplanets \citep{Wolff2017,Christiaens2019,Stolker2020,Wang2020,Benisty2021,ChoksiChiang2024}.

Hydrodynamic simulations of this late stage of planet formation have demonstrated that CPDs can form around embedded planets, but require a sufficient level of cooling in order for the gas to settle into a disk-like structure \citep{Ayliffe2009,Szul2014,Szulagyi2016,Fung2019,Marleau2023}. In accompanying work, \cite{Legaetal}, hereafter Paper 1, explored the gas flow around an embedded Jupiter-mass planet to characterise the properties of the gas as it passes from the PPD into the gravitational influence of the planet. In the so-called Nominal case (using MMSN gas densities at a distance of 5.2 AU) the gas is accreted into the planet's Hill sphere and cools too slowly to form a CPD and instead remains in a spherically symmetric cloud around the planet. If the surrounding gas density and/or the gas viscosity is reduced, the resulting gas accretion flux allows the gas to cool sufficiently to form a CPD or a rotating structure that is partially supported by pressure that we also call for simplicity CPD. In this work, we consider the simulations results of Paper 1 and explore the chemical evolution of the gas as it moves from the PPD into the CPD.

The primary goal of this work is to understand the chemical environment of the gas surrounding a young proto-planet. This helps to understand two general science questions: 1) how is the chemical complexity of the protoplanetary disk gas and dust imprinted onto the growing planet? And 2) are their unique molecular species that make good tracers of circumplanetary material - perhaps offering a unique way of detecting CPDs? We posit that the gas entering into the planet's gravitational influence from the PPD undergoes sufficient chemical processing in the hotter, denser environment of the planet's gravitational well that it evolves towards a different chemical state than is seen in the PPD. This difference may cause the rate of accretion of different volatile species to change relative to simpler approaches \citep[for ex. in][]{Cridland2020}, and may cause some molecules to appear much brighter in the CPD than in the PPD, enabling a reliable detection method.

Gas line observations have long been used to study the physical and chemical properties of the the Interstellar medium (ISM) due to their sensitivity to the local properties of the gas and ice \citep{Zuckerman1976,Draine1978,Kaufman1996,Bergin2002,vanDishoeck2021}. As we will see below, the circumplanetary material presents a significant increase in gas temperature compared to the surrounding PPD which should drive high-temperature chemistry. We expect that these higher temperatures will free frozen species - particularly water - which will further drive oxidising reactions due to the freed OH radical. For this reason we may expect that the gas around a growing planet has a lower carbon-to-oxygen ratio (C/O) than the gas in the PPD. \cite{Jiang2023} have recently argued the opposite effect of a forming planet, focusing mainly on the release of frozen CH$_4$ due to heating at a distance of a Hill radius from the planet. Here we resolve deeper into the planet's gravitational well and thus reach high enough temperatures to release all of the ice budget into the gas. The ice tends to be oxygen rich as its main constituent is H$_2$O \citep[see for ex.][]{Walsh15,Crid19b,WISHreview2021,McClure2023}.

The paper is structured as follows: in section \ref{sec:methods} we outline the methods used in this work, including the relevant details of Paper 1. In section \ref{sec:results} we present our main results of combining the numerical work of Paper 1 with the chemical model presented here. In section \ref{sec:discussion} we discuss the relevance of this work and compare to a model of the PDS 70 system. Our conclusions are presented in section \ref{sec:conclusion}.

\section{ Methods }\label{sec:methods}

In this follow up work, we post-process the numerical results of Paper 1 using the \textit{ALCHEMIST}\footnote{https://www2.mpia-hd.mpg.de/homes/semenov/index.html} \citep{Sem10,Sem17} zero dimensional astrochemistry solver to compute the chemical evolution of gas as it moves along streamlines from the PPD into the gravitational influence of the embedded planet.

\subsection{Hydrodynamics simulations}

In Paper 1 we explored different outcomes for the distribution of the accreted gas around the embedded planet.  Table~\ref{tab:Simparams} summarises the parameters of each model considered. The physical link between accretion rate and the gas distribution is through the gas cooling efficiency. For higher accretion rates (the so-called nominal model in table \ref{tab:Simparams}) the gas is not able to cool quickly enough to settle into a rotationally supported disk. Instead, it remains in a pressure supported, spherically symmetric, and hot ($\sim$600 - 1000~K) envelope. Reducing the initial surface density of the PPD at the location of the planet (LowMass model) and/or reducing the viscosity of the gas (LowMassLowVis) results in a sufficiently low accretion rate such that the gas could cool enough for a CPD to form (see also \cite{2024arXiv240214638K} for a detailed thermodynamical criterion of CPDs formation).

In this paper, we will primarily focus on the LowMass model from Paper 1, as it produced a CPD around the planet by reducing the surrounding PPD density by a factor of 10. Physically, this reduction is consistent with an older disk that has accreted a significant amount of its mass onto the host star and/or into building giant planets. We discuss the implication of this timing in a later section. Here we will briefly outline the relevant information of the numerical setup and streamline calculation, see Paper 1 for further details. We then outline the chemical setup with its primary assumptions, and outline our analysis techniques.


A common definition that we will use throughout this work is the planet's Hill radius. The Hill radius represents the size of a sphere centred on any gravitating secondary body within which the gravitational influence of the secondary exceeds that of the primary. In our case, the relevant Hill sphere is that of the embedded planet and its radius is defined as: \begin{align}
    {\rm R}_{\rm Hill} = a_p\left(\frac{M_p}{3 M_*}\right)^{\frac{1}{3}},
    \label{eq:hillradius}
\end{align}
where $a_p$ and $M_p$ is the semi-major axis of the planet's orbit and mass of the planet respectively, and $M_*$ is the mass of the host star. Our general numerical setup, presented in Paper 1, simulates a Jupiter mass planet orbiting at 5.2 AU from the Sun, such that the relevant parameters are $a_p = 5.2$ AU, $M_p = 0.000954$ $M_\odot$, and $M_* = M_\odot$, where $M_\odot$ is the solar mass. As such, the Hill sphere of the embedded planet has a size of R$_{\rm Hill} = 0.355$ AU.

\subsubsection{Numerical setup and streamlines}

\begin{figure}
    \centering
    \includegraphics[width=0.5\textwidth]{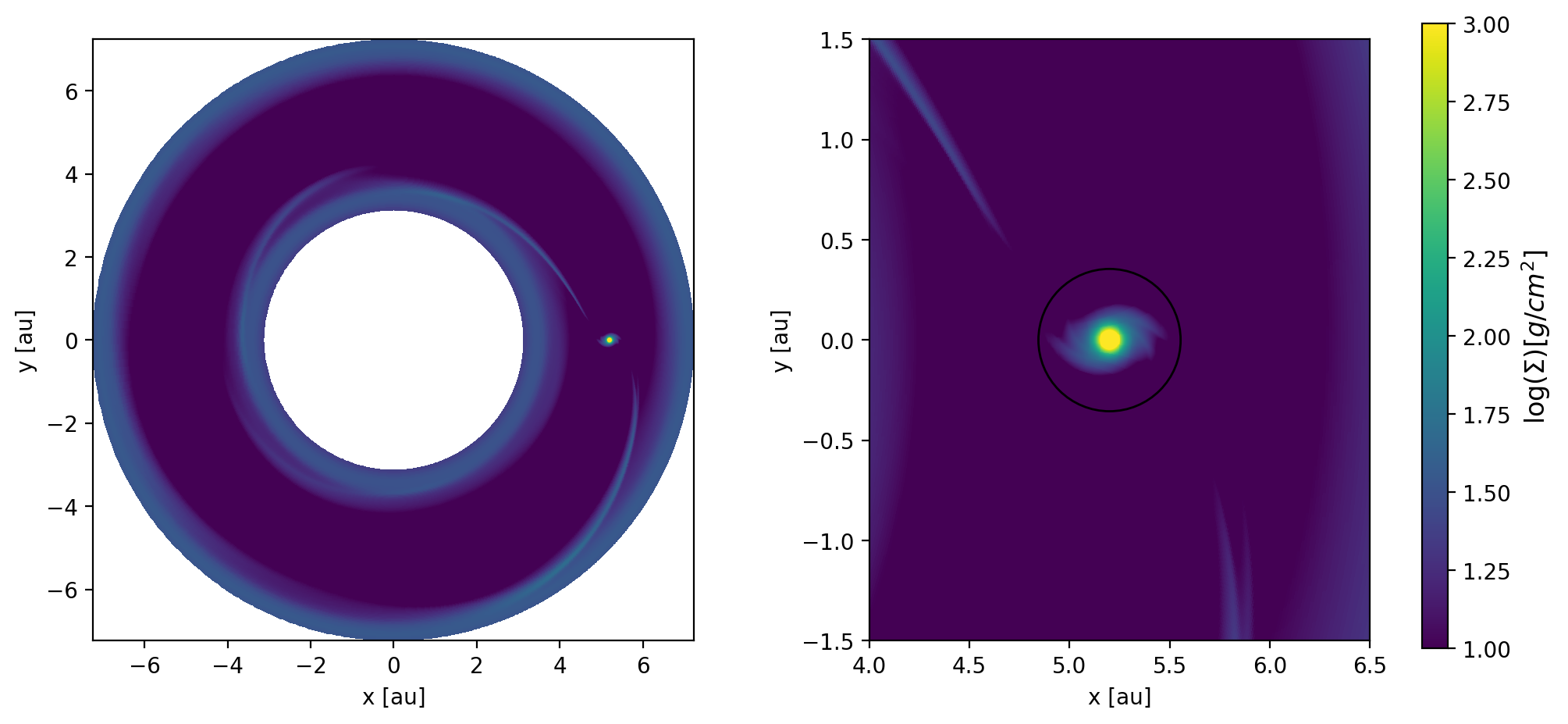}
    \caption{Example of the output from \textit{fargOCA} as seen in Paper 1. The figure axes are in units of AU while the colour range shows the surface density of the gas. The black circle shows the extent of the planet's Hill radius.}
    \label{fig:simresult}
\end{figure}

\begin{table}
    \centering
    \begin{tabular}{|l|c|c|}
    \hline
    Model name & $\Sigma_0/ M_{\sun}r_p^{-2}$ & $\nu / r^2_p\Omega_p$ \\
    \hline
    Nominal & $6.67 \times 10^{-4}$ &  $10^{-5}$ \\\hline
    \textbf{LowMass} &   $\bf 6.67 \times 10^{-5}$ &  $\bf 10^{-5}$ \\\hline
    LowMassLowVis & $6.67 \times 10^{-5}$  &  $10^{-6}$ \\\hline
    \end{tabular}
    \caption{Simulation parameters from Paper 1. The model that we will primarily focus on is in bold, although we will compute one streamline in the Nominal model.}
    \tablefoot{$\Sigma_0$ is the initial gas density surrounding the embedded planet and $\nu$ is the viscosity of the gas. $r_p$ and $\Omega_p$ are the orbital radius of the planet and the orbital frequency at that radius. M$_\odot$ is the solar mass.}
    \label{tab:Simparams}
\end{table}

Paper 1 computes the 3D flow of gas in the PPD around an embedded, Jupiter-mass planet, using the \textit{fargOCA} code \citep{2014MNRAS.440..683L} with a specific grid's implementation  for the study of details in a planet's Hill sphere introduced in \citep{LL17,LLNCM19}. The simulation domain is a full annulus extending from 3 to 7 AU and one hemisphere of the disk. The boundary conditions for the height, azimuthal, and radius are reflective, mirror, and evanescent boundaries respectively. The reference fields for the evanescent boundaries are based on the results of two dimensional, axi-symmetric simulations that reach thermal equilibrium in order to minimize wave reflections \citep{deValBorroetal2006}. The dust is not evolved and a constant dust-to-gas ratio of 0.01, is assumed in order to compute the disk's opacity. See Paper 1 for further details.

We show an example of the gas distribution from the LowMass model of Paper 1 in figure \ref{fig:simresult}. The figure shows the column density (integrated up the z-axis) of the gas over the entire computation range (left panel) as well as locally to the embedded planet (right panel). The planet cuts a deep gap, with a reduction in gas density of between a factor of 10-100 relative to the surrounding PPD gas. Both the gas flow geometry and the gap are common among all of the simulations shown in Paper 1, however the exact reduction in gas density and the shape of the gap depend on the simulated viscosity and starting density (see Paper 1).

The data that we present in this paper concern a state in which the gas flow has reached a quasi-steady state, where the velocity field does not vary significantly from one timestep to the next. In addition we have shown in Paper 1 that the energy budget of the gas near the planet evolves over timescales of thousands of orbits, implying that the temperature structure also reaches a quasi-steady state. For these reasons, and with the intention to draw streamlines through the velocity field on which we compute our chemical evolution, we take a single snapshot of the simulation to represent the physical structure of gas around the planet. The velocity field is thus kept static during the streamline integration, and the chemistry is computed on the streamline as it evolves from the PPD towards the embedded planet. Paper 1 shows multiple examples of streamlines and discusses some of their general properties. In the streamlines we assume that the dust is co-moving with the gas at a dust-to-gas ratio of $10^{-4}$, we thus assume the chemically relevant dust are composed of small grains. This is discussed in more detail in the following section.

For a better understanding of the global properties of CPDs from a finite set of streamlines, we first characterise the streamlines into a set of `families'. These families are classified by the amount of time that the streamlines spend within the CPD relative to total time of the integration, a quantity that is relevant for our chemical study. Specifically, this residence time is computed by taking the difference of $t_0$, the time when the streamline first enters within $r \le 0.4$ R$_{\rm Hill}$, i.e. in the region where gas is potentially bounded to the planet as found in Paper 1, and $t_f$ when (if) the streamline leaves the Hill sphere of the planet. If the streamline never leaves the planet's Hill sphere, then $t_f$ is set to the total integration time $t_T$. With that, the residence time is defined as:\begin{align}
    t_R = \frac{t_f - t_0}{t_T - t_0}.
    \label{eq:restime}
\end{align}
The residence time is normalised such that the time between the beginning of the simulation and the point when the streamline first crosses within 0.4 R$_{\rm Hill}$ of the embedded planet is ignored. 

\subsubsection{Treatment of dust}

There is no dust included in the simulations of Paper 1, however a constant dust-to-gas ratio is assumed in order to compute dust opacity. Dust plays a very important role in setting the chemical state of the protoplanetary disk. Physically it impacts the thermal balance of the gas by acting as both a strong opacity source to incoming radiation, as well as the primary cooling source. Chemically, dust grains provide the freeze-out site for volatile molecular species when the gas and dust temperatures are sufficient cold, and their surface acts as a catalyst for some chemical reactions. 

For the purpose of our chemical calculation we assume that the disk is sufficiently old (given that it contains an embedded giant planet) so that most of the dust mass has grown and settled to the midplane of the protoplanetary disk. Thus the majority of the dust mass will be decoupled from the gas flow and be less important to further chemical evolution as the gas flows from the PPD towards the embedded planet. In the chemical calculations that we will discuss below, we assume that the (chemically) relevant dust grains are strictly small, and dynamically coupled to the gas. From this assumption, the dust density is proportional to the gas density, with a constant (in both space and time) dust-to-gas mass ratio of $10^{-4}$. This low ratio, 100 times smaller than is seen in the ISM, is typical for the small grain population in dynamic dust evolution models that include growth and fragmentation in steady state \citep{DD05,B10}. This low ratio is often observed in the upper atmospheres of protoplanetary disks after the majority of the dust mass has settled \citep{Dal2006,Meijerink2009,Bruderer2013,Bosman2018}. 

While the small dust grains make up a very small proportion of the dust mass in the disk, their total number dominate over the large grains \citep[based on the MRN distribution,][]{MRN77}. Through this one can find that the small grain population dominates the total dust surface area in the disk and is thus the most important population of grains when it comes to chemical evolution.

\begin{figure*}
    \centering
    \begin{minipage}{0.33\textwidth}
        \includegraphics[width=\textwidth]{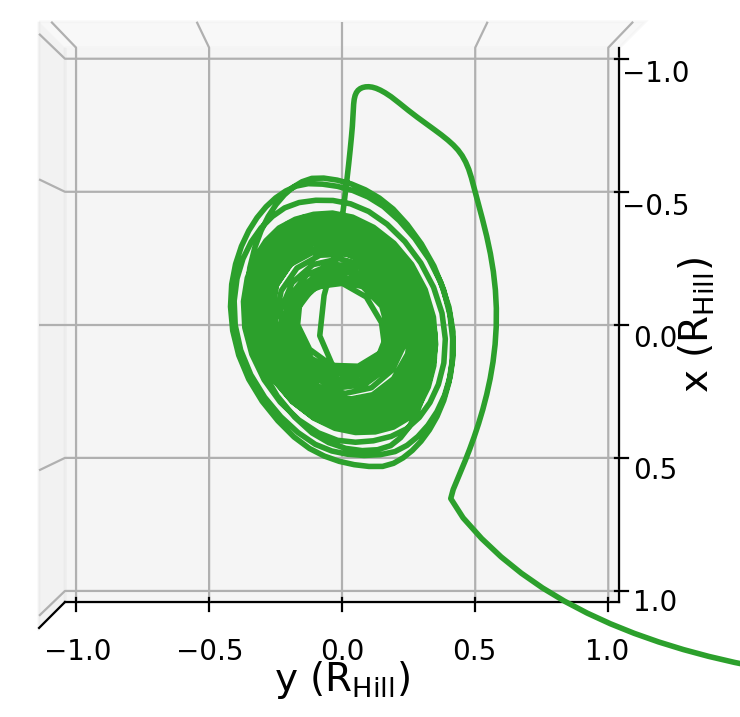}
    \end{minipage}
    \begin{minipage}{0.005\textwidth}
        \hfill
    \end{minipage}
    \begin{minipage}{0.33\textwidth}
        \includegraphics[width=\textwidth]{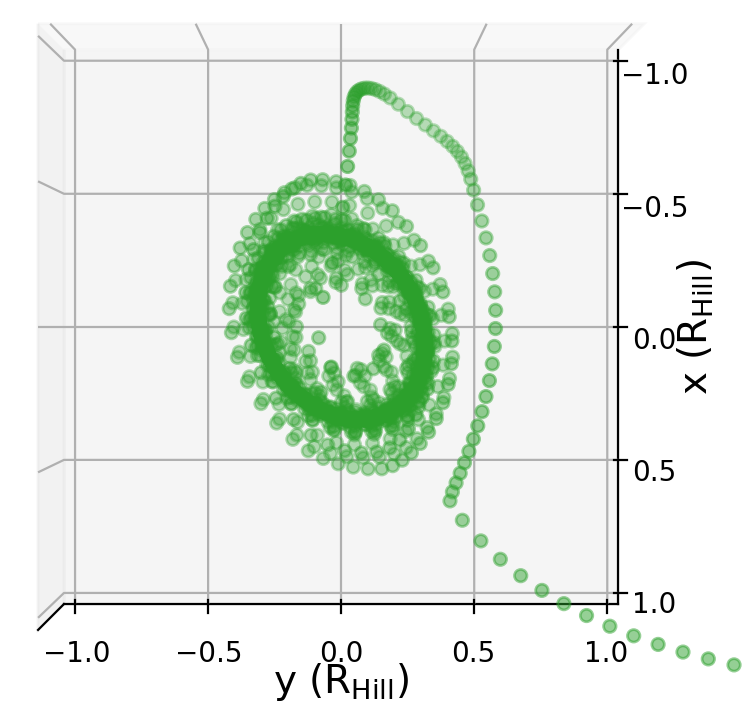}
    \end{minipage}
    \begin{minipage}{0.005\textwidth}
        \hfill
    \end{minipage}
    \begin{minipage}{0.33\textwidth}
        \includegraphics[width=\textwidth]{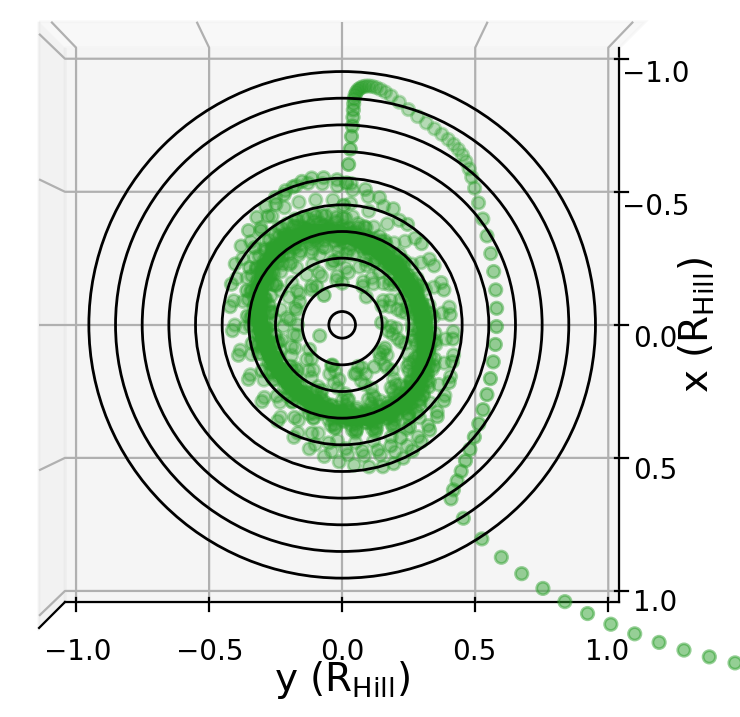}
    \end{minipage}
    \caption{Visualisation of the griding for the column number density calculation, viewing down along the z-axis. The left panel shows a zoom into a streamline that enters into the CPD and remains for the bulk of the integration. The axes show distance relative to the location of the planet, in a rotating reference frame. The middle panel is the same as the left panel, but shows the points along the streamline computed during the integration. The right panel shows each individual point computed along the streamline as well as circles on which we compute the average volume number density of a given molecule at a given $r$ and $z$ in cylindrical coordinates.}
    \label{fig:column_calc}
\end{figure*}

\subsection{ Chemical model }

\subsubsection{Physical setup}

We use the zero dimensional, time dependent astrochemistry code \textit{ALCHEMIC} \citep{Sem10,Sem17} to solve the chemical evolution of the gas. For each position along the streamline for which we compute chemistry, we extract the temperature and volume density of the gas from the simulation grid of Paper 1 using the \textsc{RegularGridInterpolator} package of \textsc{Scipy} after interpolating the temperature and density values from the irregular grid produced by \textit{fargOCA} onto a regular cartesian grid using the \textsc{Interpolate} package of \textsc{Scipy}. The interpolation is done linearly.

At each point along the streamline the column density of hydrogen gas is computed towards both the host star as well as vertically upwards from the midplane. This is done in order to compute the relative importance of high energy photons coming from both the host star and from the surrounding ISM. We assume that the host star is a standard T Tauri star with a standard UV field and X-ray luminosity of $10^{31}$ erg/s. We find, however that the column density towards the host star implies very a large extinction ($A_v\sim 10^4$) after using a relation between the H$_2$ column density and optical extinction observed by \cite{TolgaOzel2009}. On the other hand the minimum vertical extinction is on the order of $A_v\sim 4-10$ so that the ISM UV field is important for the chemical evolution. For this we use the UV radiation field of \cite{Draine1978} with the extensions made by \cite{vDB1982}, and assume $G_0 = 1$ at the top of the disk.

We follow the gas as it moves along streamlines from a chosen initial condition. We assume that the gas and dust in the streamline does not interact with the surrounding material, however its temperature and density evolves as the streamline moves through the simulation domain. In this way we ignore diffusion and mixing with the surrounding gas but include possible changes in chemical properties due to changes in the freeze out, desorption, and chemical reaction rates due to changes in gas density and temperature. We assume that the gas velocity field is static over the whole integration time, about 80 orbits in total. From a single initial position, chosen near the planet's Hill sphere, we integrate the streamline in both the forward (in time) direction as well as the backward direction.

As the streamline evolves forward in time it moves further into the gravitational potential of the embedded planet, heating up as it does so. A subset of streamlines will pass close to the embedded planet but will not be capture by its gravitational potential and instead will return into the protoplanetary disk. A second group will move into orbit around the embedded planet, representing the formed CPD. We explore the difference between these groups below. The backward integration evolves the streamline away from the embedded planet further into the protoplanetary disk. There, the streamline encounters features of the disk such as the edge of the planet gap and the spiral arms induced by the embedded planet (note figure \ref{fig:simresult}) where the density and temperature change slightly compared to the background gas distribution of the protoplanetary disk (see also Paper 1 for a detailed description of streamline behaviour). 

\subsubsection{Chemical network}

The \textit{ALCHEMIC} code is equipped with the \verb|osu_03_2008| rate file and chemical network with extensions based on the Kinetic Database for Astrochemistry \citep[KIDA,][]{Wakelam2012}. The network includes 6065 chemical reactions of 655 molecular and atomic species. The network includes primarily two-body gas phase reactions such as ion-molecule and neutral-neutral interactions, cosmic ray ionization and dissociation, gas-grain interactions including neutral and ion absorption as well as thermal-, photo-, and cosmic ray driven-desorption. Finally, the network includes dust grain surface reactions which allow absorbed species to move along the grain lattice and react with other species. The network was featured in \cite{Semenov2018}, see that work for more detail.

For the main results of this work we use the network that accompanied ALCHEMIC for computing the chemical evolution of the gas as it moves through the streamline from the PPD to the CPD. Later we will compare our results to separate chemical models of the PDS 70 protoplanetary disk presented in \cite{Crid2023}. These models were run using the DALI thermochemical code \citep{Bruderer2012,Bruderer2013} using the chemical network of \cite{Miotello2019} which is based on \cite{Visser2018}. This network is largely based on the \verb|osu_03_2008| network, but has been optimised to compute C, O, and N chemistry more (computationally) efficient, meaning that some chemical species - like sulfur - have only limited reactions in the network. We will discuss this later in the paper.

\subsubsection{Chemical evolution}

The chemical evolution proceeds as follows: the initial temperature and density of the gas is set at the final position of the backward integrating streamline (the coloured triangles in figure \ref{fig:cstreams}). The chemistry at this position is initialised with the abundances shown in table \ref{tab:Initabn} and integrated for 1 million years (Myr) so that the chemical state of the gas clump reaches a steady state before we allow it to begin moving through the disk. Waiting for 1 Myr to pass is meant to replicate the fact that the protoplanetary disk is at least this old by the time the giant planet has reached the evolutionary phase in which we have initialised the hydrodynamic simulation. The initial Myr also allows the elemental initial conditions time to evolve towards molecular species like H$_2$O, CO, CO$_2$, and H$_2$S that are relevant in PPDs.

After 1 Myr of evolution we `release' the clump and follow its evolution across the velocity field, we consider $t=0$ as the point where the clump is released. The timestep $dt$ of the streamline evolution is set to 0.16$\Omega_p^{-1}$, or about a sixth of an orbital period. For a given $t\ge 0$ the physical properties of the gas are interpolated as described above for the position along the streamline at $t$. The chemistry is then integrated to a time of $t+dt$ with a constant temperature and density. The temperature and density are then updated to their interpolated value for a position at time $t+dt$ along the streamline before the chemical evolution is continued.

To facilitate the above algorithm we slightly modified the \textit{ALCHEMIC} code so that the resulting chemical state computed at one timestep is used as the initial condition for the next timestep. The coupled evolution of the disk chemistry and the physical properties of the disk is similar to the method of \cite{Walsh15,Eistrup2016}, and matches the modification made in \cite{Crid19c}. 

\begin{table}[]
    \centering
    \begin{tabular}{|l|c|l|c|}
    \hline
    Species & Abundance & Species & Abundance \\\hline
        H$_2$ & 0.495 & Si & 7.940 $\times 10^{-8}$\\\hline
        H & 0.01 & Na & 1.738 $\times 10^{-8}$\\\hline
        He & 0.851 $\times 10^{-1}$ & Mg & 1.000 $\times 10^{-11}$\\\hline
        C & 2.692 $\times 10^{-4}$ & Fe & 4.270 $\times 10^{-9}$\\\hline
        N & 5.350 $\times 10^{-6}$ & P & 2.570 $\times 10^{-9}$\\\hline
        O & 2.880 $\times 10^{-4}$ & Cl & 3.162 $\times 10^{-9}$\\\hline
        S & 1.910 $\times 10^{-8}$ & f$_{\rm dtg}$ & $10^{-4}$ \\\hline
        $\chi_{\rm CR}$ & $10^{-17}$ s$^{-1}$ && \\\hline
    \end{tabular}
    \caption{The initial abundances, relative to the total number of hydrogen atoms ($2n_{\rm H_2} + n_H$). These values are taken from \cite{Bosman2021MAPS}, although we ignore the CO depletion that they assumed. The abundances of He, P, and Cl are taken from the default \textit{ALCHEMIC} chemical network. The heavier elements are mainly important for supplying ions into the chemical system. The dust-to-gas mass ratio (f$_{\rm dtg}$) and the selected cosmic ray ionization rate ($\chi_{\rm CR}$) is also included.}
    \label{tab:Initabn}
\end{table}

\subsubsection{Initial elemental abundances}

\begin{figure*}
    \centering
    \includegraphics[width=0.8\textwidth]{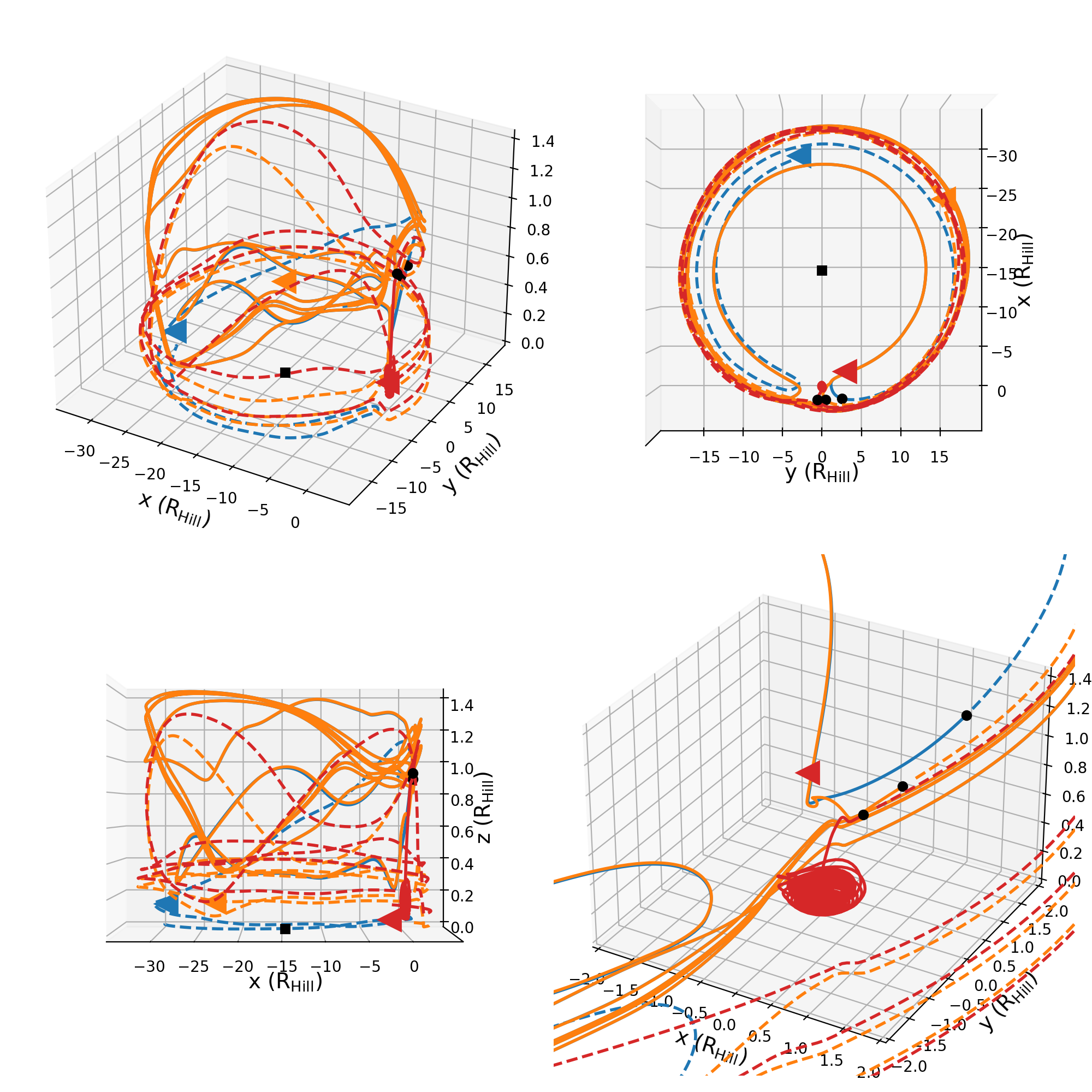}
    \caption{Example of some of the streamlines by the embedded planet. Three types of streamlines are shown: one misses the CPD completely [(1) blue], one enters into the region of the CPD for less than an orbit [(2) orange], and one that ends up in the CPD [(3) red]. The solid lines represent forward integration from the initial conditions (black circles), while the dashed lines represent backward integration from the initial conditions. The coloured triangles show the location from which the chemical calculation is initialized. The host star is the black square. The x-axis corresponds to the radial distance along the axis passing through the host star and the embedded planet, centred here on the embedded planet. The y-axis runs perpendicular to the host star - planet axis and the z-axis extends from the midplane to heights above the disk. $x > 0$ corresponds to distance farther from the host star than the planet, while $y>0$ corresponds to the leading side of the planet's orbit.}
    \label{fig:cstreams}
\end{figure*}

The initial abundances are shown in table \ref{tab:Initabn}, and are used to initialise the first 1 Myr of chemical evolution. The position of the gas when it is being initialized in this way is shown by the coloured triangles in figure \ref{fig:cstreams}. The majority of elements are initialised as atoms, apart from H$_2$, and evolve to form the molecular species in the PPD. The most abundant molecular species to be produced during this initial 1 Myr of evolution are shown in figure \ref{fig:initabun}. \cite{Eistrup2018} showed that there are differences in the outcome of chemical calculations depending on whether the system is initialised by elemental or molecular abundances (known as the `reset' and `inheritance' scenarios respectively). For our work, this distinction is less important because the gas flows into regions of the disk that are sufficiently warm to effectively reset much of the chemical evolution that occurred prior. 

We derived our initial elemental abundances from \cite{Bosman2021MAPS}, however we ignore the `CO depletion' that they model which adjusts the available carbon and oxygen to further chemical evolution. This CO depletion is meant to replicate the freeze out of volatiles and settling of dust grains that can remove volatile elements from the molecular layer of protoplanetary disks \citep[see for ex.][]{Krijt2020}. While we ignore this CO depletion, we include a reduction in the dust-to-gas ratio, assuming that the only chemically relevant dust grains are the smallest ones. We thus possibly overestimating the total number of carbon and oxygen atoms available to the chemical evolution of the gas in the streamlines.

\subsection{Estimating CPD observations}

One of our goals in this work is to understand whether there are observational tracers that are uniquely tied to the presence of circumplanetary material. One thing we will find is that the molecular species that are most abundant (i.e. have optically thick emission), like CO, are poor tracers because differentiating between the PPD and CPD is difficult. On the other hand, optically thin tracers - particularly rarer species that are not abundant in the PPD may turn out to be better tracers of the warm circumplanetary material. As such, their emission will most depend on the total column of a given species along the line of sight. We will estimate the column number density of the molecules that we are interested in, however for simplicity we do not compute the astrochemistry across the whole numerical regime. Instead, we use the positions and chemical properties of the streamline that occupies the CPD to estimate the abundance distributions of molecules around the embedded planet. 

The calculation of the column number density proceeds as follows: we focus only on the space centred on the planet and extending out to 1 Hill radius in all directions. For a streamline that orbits in the CPD we have a large set of coordinates (x,y,z) with corresponding temperature, density, and chemical abundance of the gas at those points along the streamline. Over a range of radial distances, $r$, from the planet we average the abundances of a given molecule over the azimuth amongst the nearest neighbouring coordinates around the circle of radius $r$. This is done over a large set of heights above the midplane of the CPD and the total column is integrated towards the observer who, we assume, is viewing the disk face on.

In figure \ref{fig:column_calc} we visualise how the streamlines are used to estimate the CPD column. Here we show a streamline that enters and stays in the CPD (left) as well as the individual points that are computed by the integration routines of Paper 1 (middle). Finally in the right panel we show the set of concentric circles centred around the embedded planet on which we compute the average volume density of a molecule. For 10 separate heights on and above the midplane of the CPD we estimate the average volume density of a given molecule, and integrate along the z-axis. The reported column number density below is over a length scale of 1 Hill radius around the embedded planet, corresponding to a physical lengths scale of 0.355 AU. The results of these calculations are discussed in a later section.

\begin{figure*}
    \centering
    \includegraphics[width=0.8\textwidth]{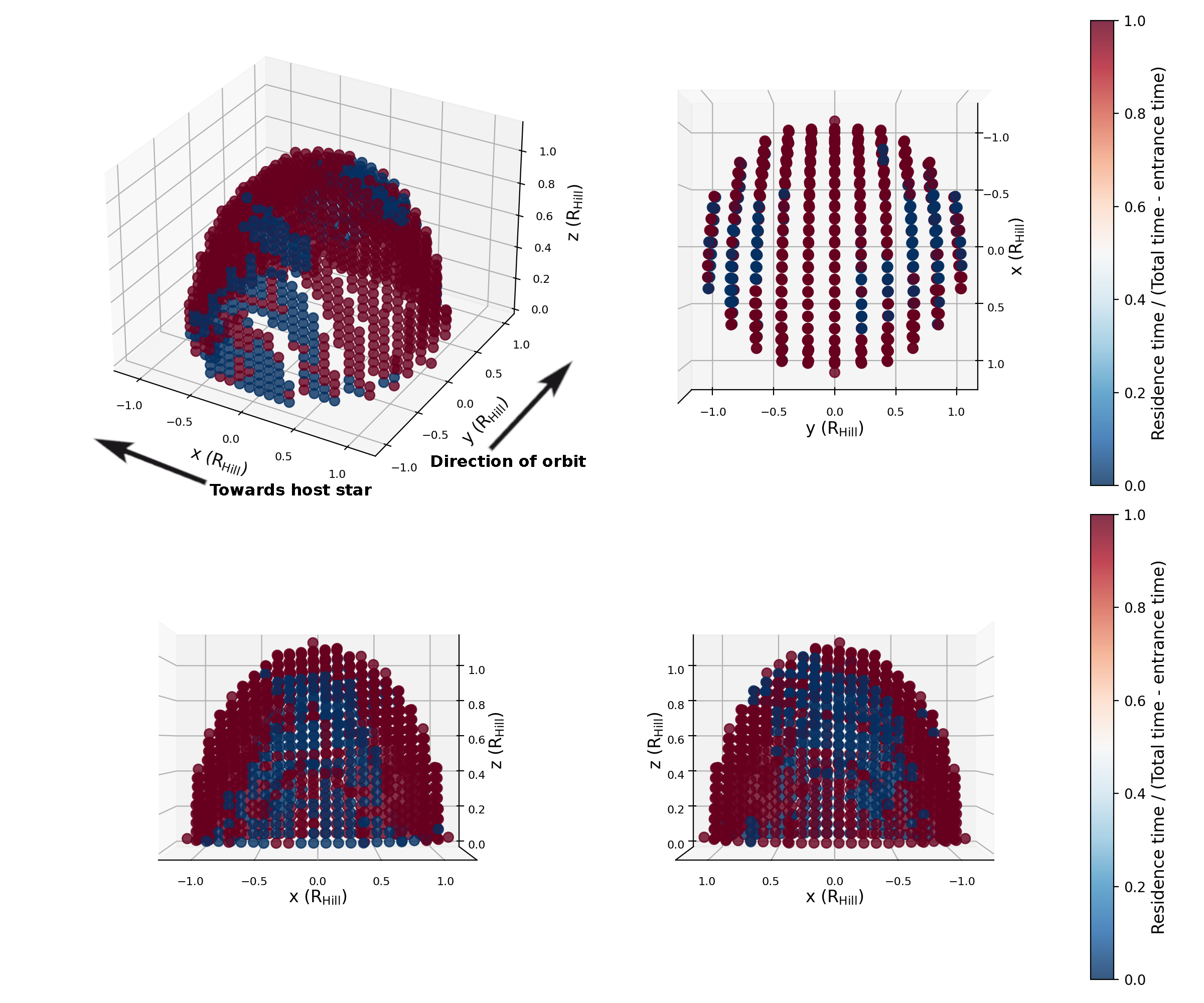}
    \caption{Map of the residence times for a set of initial conditions begin at a radial distance of 1 R$_{\rm Hill}$ from the embedded planet. Each point represents an individual initial condition for the streamline (forward) integration and is coloured by their residence time relative to the difference between the total integration time and the time that the streamline entered into the CPD. The x-axis corresponds to the radial distance along the axis passing through the host star and the embedded planet, centred here on the embedded planet. The y-axis runs perpendicular to the host star - planet axis and the z-axis extends from the midplane to heights above the disk. $x > 0$ corresponds to distance farther from the host star than the planet, while $y>0$ corresponds to the leading side of the planet's orbit. Each panel represents the same calculation, but shows different rotations in cartesian space. The bottom left panel represents strictly trailing side initial conditions ($y<0$) while the bottom right panels shows the leading side initial conditions ($y>0$).}
    \label{fig:residence_map}
\end{figure*}

\section{Results}\label{sec:results}

\subsection{CPD streamline properties}


To understand the chemical changes in the gas caused by the local heating in the CPD, we select a subset of streamlines, with different orbital history and study the temporal evolution of the gas as it passes into CPD and any lasting effects as the gas leaves the CPD. 
Most of our streamlines are initialised near the embedded planet, above the midplane (black circles in figure \ref{fig:cstreams}). Some of these streamlines thus represent gas flow that is commonly called meridional flow. A common evolutionary trend in our numerical simulation is that these streamlines enter into the CPD through this meridional flow, and remain in orbit around the embedded planet. There are a few streamlines that approach near ($< 0.4$ R$_{\rm{Hill}}$) the embedded planet, but soon after exits the CPD from either the first or second Lagrange point. Gas exiting via the first Lagrange point then co-orbits with the planet at an orbital radius slightly smaller than the planet's orbital radius while gas exiting the second Lagrange point co-orbits at a slightly larger radius.

To better generalise the chemical properties of the CPD we categorise streamlines into 3 families. Each family is characterised by the amount of time spent in the vicinity of the embedded planet as defined in Eq.\ref{eq:restime}. As shown in Paper 1, in the LowMass case, the hot inner envelope extends up to $0.2-0.3$ Hill radii while orbits captured in the CPD structure have distances from the planet in the interval $0.3-0.4 R_H$. The residence time ($t_R$) measures the amount of time that passes between the gas passing inward of 0.4~R$_{\rm Hill}$ from the embedded planet and the gas passing a distance of 1 R$_{\rm Hill}$ from the planet. The limiting distance of $0.4 ~R_{\rm Hill}$ comes from Paper 1, since gas can be bounded to the planet below this distance. Based on these residence times the three families are named as follows:\begin{enumerate}
    \item \textbf{missed}: $t_R=0$; a streamline that never approaches the embedded planet closer than 0.4 R$_{\rm Hill}$, blue line in figure \ref{fig:cstreams}.
    \item \textbf{escape}: $0 < t_R \le 0.9$; a streamline that approaches within 0.4 R$_{\rm Hill}$ but does not end up gravitationally bound to the embedded planet, orange line in figure \ref{fig:cstreams}.
    \item \textbf{cpd}: $t_R > 0.9$; a streamline that falls from the PPD and remains bound in the CPD for the remainder of the simulation, red line in figure \ref{fig:cstreams}.
\end{enumerate}

\begin{figure}
    \centering
    \includegraphics[width=0.5\textwidth]{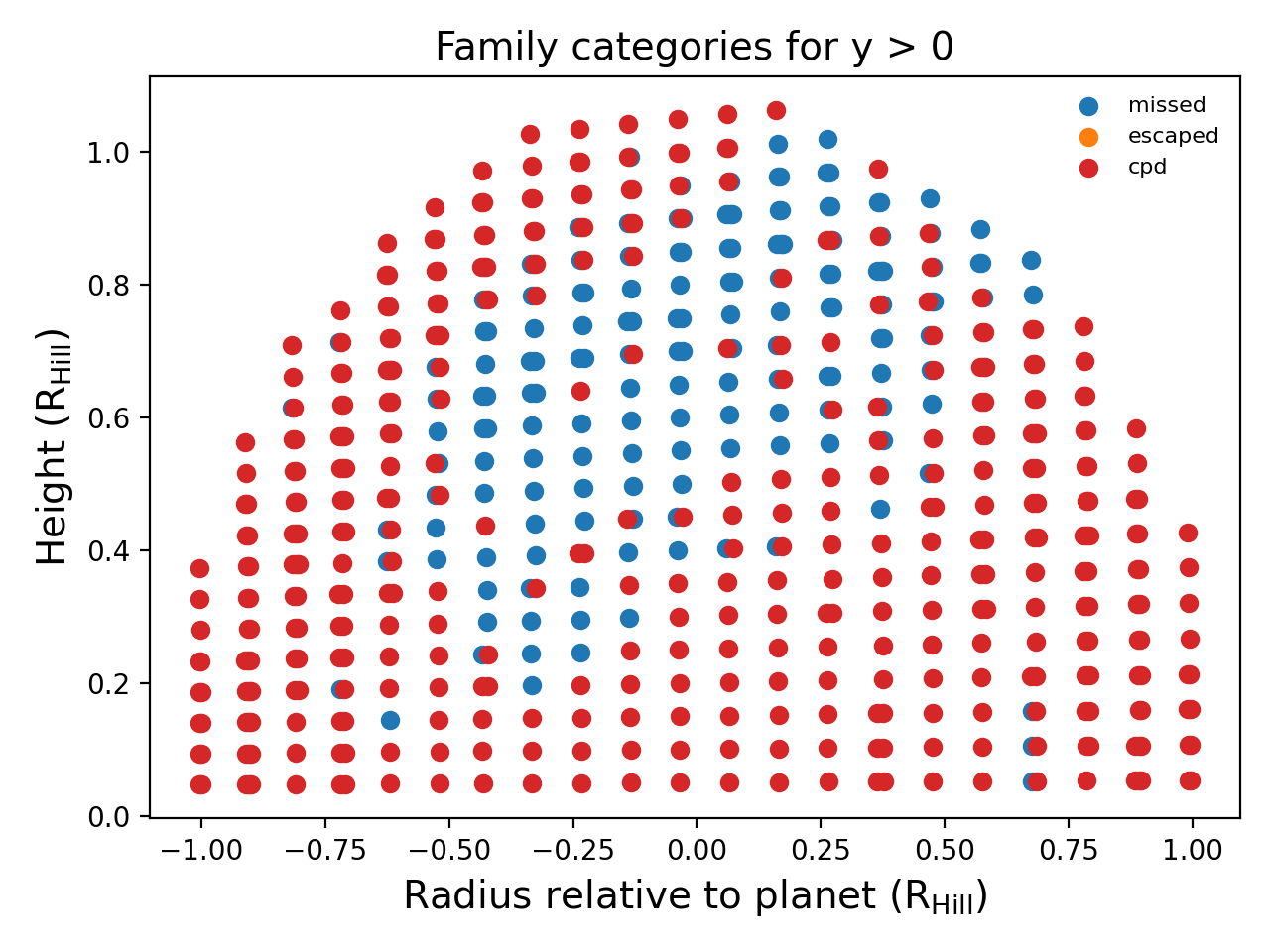}
    \caption{Families: missed (blue), escape (orange), and cpd (red) for the same set of initial conditions as in figure \ref{fig:residence_map} and $y>0$. In this particular configuration no escape initial conditions can be seen. The radius relative to the planet has the same definition as the x-axis from figure \ref{fig:residence_map}.}
    \label{fig:families}
\end{figure}

In figure \ref{fig:residence_map} we show a series of initial conditions selected on the shell centred on the embedded planet with radii spanning between 0.9 and 1 R$_{\rm Hill}$. Each coloured point on the figure represents a separate initial condition in the streamline calculation and the colour denotes the residence time (as defined above) for each streamline. On the shell of radius of 1 R$_{\rm Hill}$ there are small windows of very low residence times distributed between regions of very high residence time streamlines. Figure \ref{fig:families} shows the family representation of figure \ref{fig:residence_map}, where each coloured point represents the initial conditions sorted into each family. The plot shows the set of initial conditions along the central star - planet axis (i.e. the planet is located at the origin of the plot). In this arrangement no streamline initial conditions are found to result in escaping streamlines, which require very particular trajectories.

As a way of assessing the overall importance of each family to the global picture we compute the total gas flux entering into the CPD through each family. On a spherical shell of radius 1 R$_{\rm Hill}$ we compute the evolution of a large (10000) set of streamlines and their associated residence times. The fractional contribution to the total flux through the sphere of 1 Hill radius of each of the streamline families are: (1) missed: 46.0\%, (2) escape: 1.0\%, and (3) cpd: 53.0\%. From this we see that for the flux of gas that crosses 1 R$_{\rm Hill}$ from the embedded planet, nearly half of the gas never approaches near the CPD, while just over 50\% enters the CPD from 1 R$_{\rm Hill}$. Another tiny fraction of gas enters the region near the embedded planet for a short amount of time before returning to the PPD.

\begin{table}[]
    \centering
    \begin{tabular}{|c|c|c|c|c|}
        \hline
        Family & ${\rm R}_{\rm init}/{\rm R}_{\rm Hill}$ & $\theta_{\rm init}$ (rad) & z$_{\rm init}$/R$_{\rm Hill}$ & r$_{\rm min}$/R$_{\rm Hill}$ \\\hline
        missed & 15.67 & 0.1587 & 0.91 & 1.17 \\\hline
        escape & 15.70 & 0.0317 & 0.90 & 0.27 \\\hline
        cpd & 15.71 & -0.0317 & 0.94 & 0.02 \\\hline
        \textit{nominal} & 14.58 & 0.0 & 1.21 & 0.11 \\\hline
    \end{tabular}
    \caption{Properties of the 3 streamlines in the LowMass model as well as an example streamline from the Nominal model of Paper 1. This streamline is used for a comparison between the Nominal and LowMass models in section \ref{sec:nominal}.}
    \tablefoot{Starting coordinates of the 3 test streamlines for each family that passes near the CPD. The radial coordinate is shown relative to the position of the host star. The embedded planet orbits at R$_0 = 5.2$ AU $=14.6$ R$_{\rm Hill}$. r$_{\rm min}$ refers to the minimum approach of each streamline to the embedded planet.}
    \label{tab:initcon}
\end{table}

\begin{figure}
    \centering
    \includegraphics[width=0.5\textwidth]{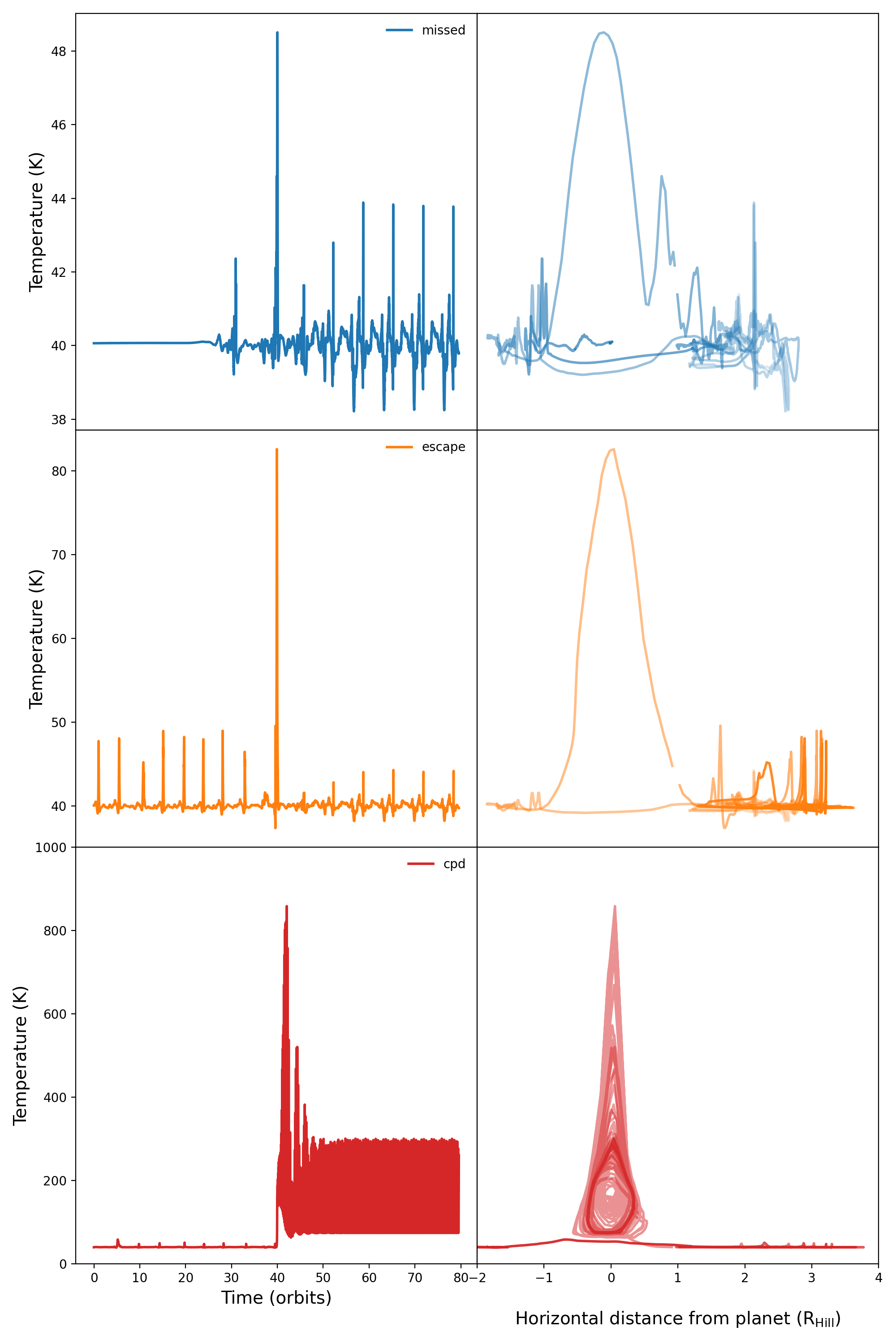}
    \caption{Temperature evolution for test streamlines. The first column shows the temperature as a function of time in orbits of the embedded planet. The second column shows the same temperature evolution of the streamlines as function of distance from the planet. The streamlines that enter the CPD first do so at around the same time ($\sim40$ orbits) because their forward integration is started just before they enter the CPD. In the second row the opacity of the line for $t>40$ orbits is reduced as a function of time to better differentiate separate orbits in the CPD.}
    \label{fig:temp}
\end{figure}

To act as representatives for their family we select one streamline each with initial conditions near the axis connecting the host star and planet, and at a vertical height near 1 R$_{\rm Hill}$. The initial positions of each test streamline is shown in table \ref{tab:initcon}. In the first column of figure~\ref{fig:temp} shows the temperature evolution for the selected streamlines for each family as a function of time. The small, early fluctuations represent the time when the streams are evolving in the protoplanetary disk, while the large fluctuations (up to 800~K) represent the streamlines closest approach to the embedded planet, where their gravitational potential energy is converted to heat. The streamlines that are caught in the CPD enter closest to the embedded planet before slowing moving outwards to lower temperatures. Their orbits appear to be eccentric because of the temperature fluctuations. The two streamlines that enter the CPD (including the escaping one) do so at nearly the same time ($\sim 40$~orbits) because their forward (and backward) integration begin at a point just prior to the gas entering the CPD. So evolution prior to 40 orbits is represented by the backward integration, while the evolution after 40 orbits is computed by the forward integration. For completeness we show the evolution of the gas density, which plays an important role in the gas chemistry, in figure \ref{fig:dens}.

As an alternative view, the second column of figure \ref{fig:temp} shows the evolution of the temperature with respect to the streamlines' distance to the embedded planet. The highest temperatures occur when the gas passes by its closest approach to the planet, when it is deepest in the planet's gravitational well, these closest approaches (r$_{\rm min}$) are shown in table \ref{tab:initcon}. The maximum temperature that each streamline reaches is directly related to the closet approach of the gas. Far from the planet, the gas fluctuates in temperature between $\sim 38$K and $\sim 44$K as it orbits in the PPD.

\subsection{Chemical evolution in the streamlines}

\subsubsection{Initial molecular abundances at the beginning of the streamline}

\begin{figure*}
    \centering
    \includegraphics[width=0.8\textwidth]{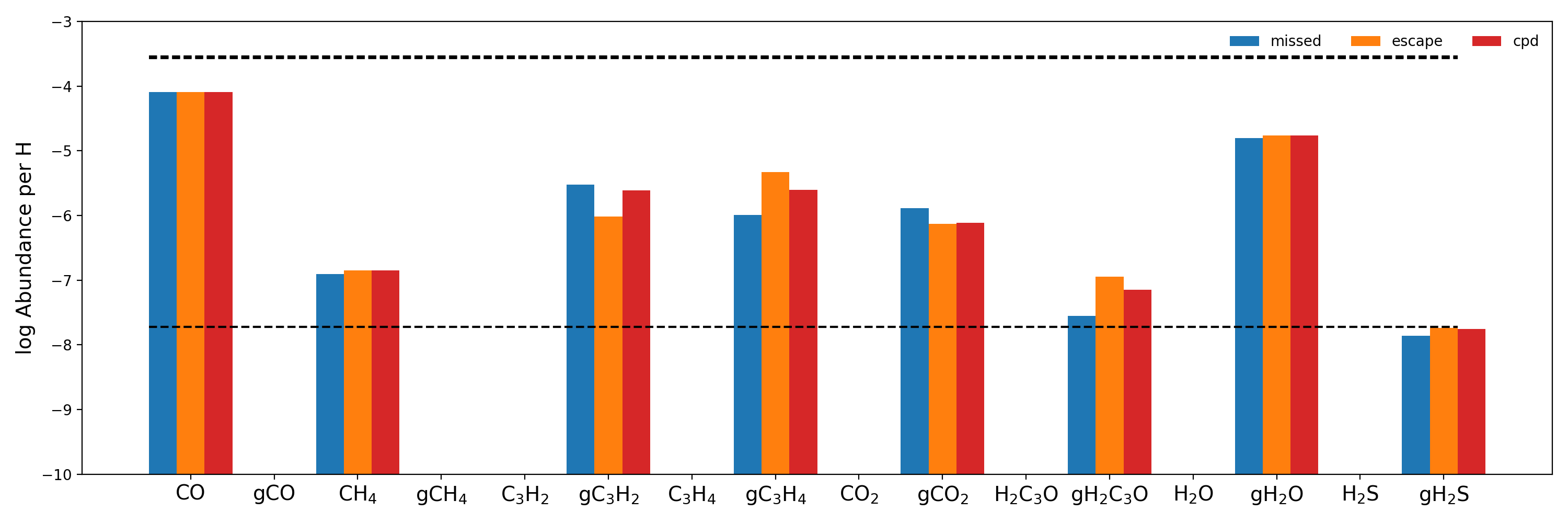}
    \caption{The most abundant initial molecules in their gas and ice phases (prefaced with a `g') for each of the streamline calculations. Apart from molecular hydrogen and helium the only other gas phase species are CO and CH$_4$. The other abundant initial species are all in the ice phase because the streamlines each start near the cold midplane of the PPD. The three dashed lines (two of which are overlapping at the top of the figure) ordered from top to bottom note the elemental abundance of oxygen, carbon, and sulfur, respectively.}
    \label{fig:initabun}
\end{figure*}

In figure \ref{fig:initabun} we show a collection of the initial molecular abundances of the most abundant species after the first 1 Myr of chemical evolution (i.e. before the gas is released in the streamline). The must abundant gaseous species with elements heavier than helium are CO and CH$_4$ since each of the streamlines begin at temperatures sufficiently high to keep these species in their gas phase. On the other side the gas is sufficiently cold that volatile species like CO$_2$ and H$_2$O are frozen onto the dust grains. Similarly there are a collection of hydrocarbons, the most abundant of which are C$_3$H$_2$ and C$_3$H$_4$, that are produced on the dust grains through successive hydrogenation of elemental carbon.

Carbon is distributed among many different species, the most abundant of which is CO gas. However there is a significant quantity of carbon distributed among many other species, mainly in the ice phase, primarily hydrocarbons of varying lengths and CO$_2$. Oxygen is mainly found in CO gas, and CO$_2$ and H$_2$O ices, with a very small quantity of organic species like H$_2$C$_3$O. Finally, sulfur can be found nearly exclusively in H$_2$S in the ice phase. This follows from the fact that the streamlines originate from the cold midplane of the PPD where H$_2$S can be produced through the hydrogenation of elemental sulfur on the dust grains.

From this initial point we will see that the changing physical properties (density, temperature) of the {gas as it travels from the PPD to the CPD drives the molecular inventory of the gas and ice species away from the initial abundances. In the following section, and with their accompanying figures we see that the molecular inventory of the CPD is driven towards molecular species that differ from what was computed in the PPD during the first 1 Myr of chemical evolution.

\begin{figure*}
    \centering
    \includegraphics[width=0.8\textwidth]{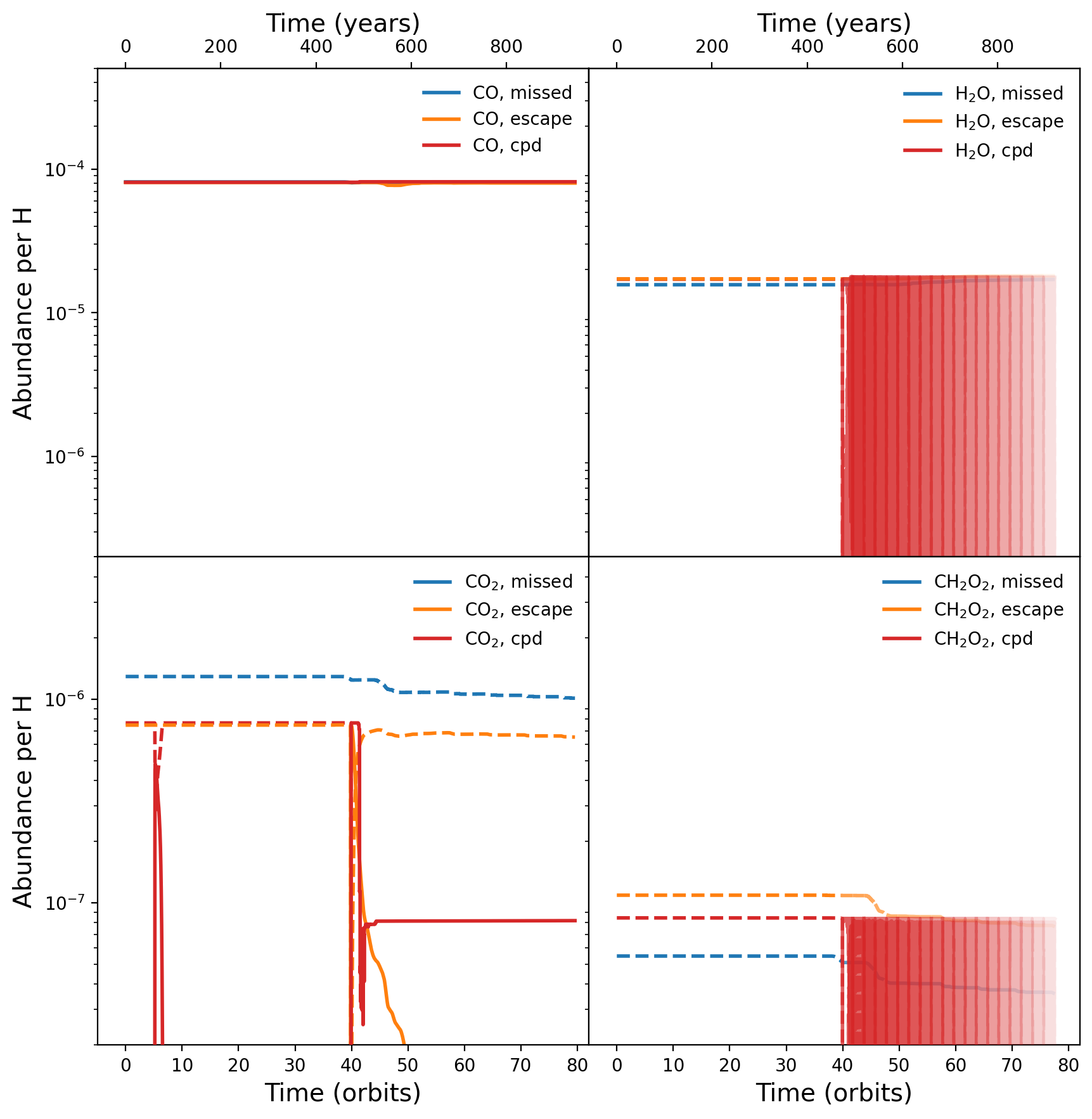}
    \caption{Evolution of the most abundant oxygen bearing species for the 3 test streamlines. The solid lines represent the gas phase of each species while the dashed lines represent the ice species. CH$_2$O$_2$ represents the most abundant `organic' species in the CPD. Note the change in y-axis range between the top and bottom rows. For water and CH$_2$O$_2$ the opacity of the line for $t>40$ orbits is slowly reduced to better see the other lines on the plot.}
    \label{fig:Obearing}
\end{figure*}

\begin{figure*}
    \centering
    \includegraphics[width=0.8\textwidth]{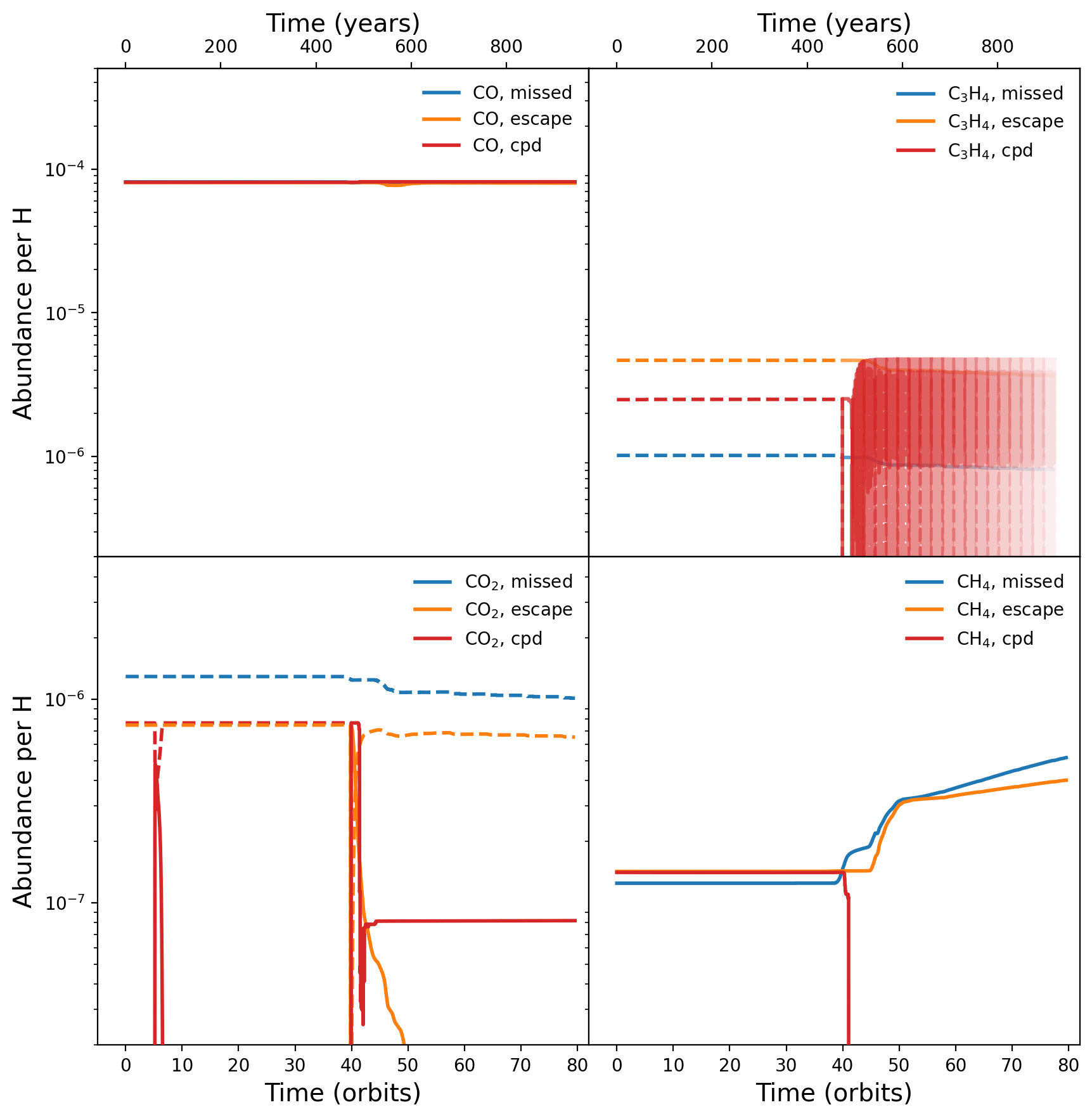}
    \caption{Same as in figure \ref{fig:Obearing} but for the most abundant carbon bearing species. Again, note the change in y-axis range between the top and bottom rows. For C$_3$H$_4$ the opacity of the line for $t>40$ orbits is slowly reduced to better see the other lines on the plot.}
    \label{fig:Cbearing}
\end{figure*}

\subsubsection{Streamline chemistry: O-bearing species}

In figure \ref{fig:Obearing} we show the most abundant oxygen bearing species in the gas of the three test streamlines. We repeat here that the streamlines enter or pass near (in the case of the flyby) the CPD near a time of 40 orbits. At early times (t$<$40 orbits) the streamlines represent the upper atmosphere of the PPD, where the gas is more strongly irradiated but colder. There, the gas reaches a chemical steady state with abundances shown in figure \ref{fig:initabun}, and only sees slight changes in abundances as the gas orbits around the PPD. As the gas enters into the gravitational influence of the planet (at t$>$40 orbits) gas species, like CO are unaffected. Ice species, on the other hand see a strong change in their abundance due to the rapid increase in gas temperature.

The most abundant ice, water, rapidly desorbs as the gas and dust temperatures (which we assume are coupled) increase. As the disk streamline (red line) enters into an orbit around the embedded planet, we see a large fluctuation in the water abundance (top right panel) between the gas and ice phases. This fluctuation is driven by the temperature fluctuation of the gas as the streamline orbits around the embedded planet (recall figure \ref{fig:temp}). The temperature fluctuation happens to span the freezing temperature of water, and thus it can continually cycles between the two phases in the isolated streamline. In figure \ref{fig:zoom} we focus on a few orbits of time evolution to better visualize the oscillation between the different phases of species like H$_2$O.

The bottom left panel shows another abundant ice in the PPD - CO$_2$ - which drops its abundance as the streamline enters the CPD, but slightly recovers in the gas phase within the CPD. In PPD chemistry the production of CO$_2$ is largely attributed to the oxidation of CO in the either the gas (at high temperatures) or the ice phase with free OH\citep{Walsh15,Eistrup2018}. H$_2$O is thermally dissociated in the gas phase which provides the OH needed to reproduce gaseous CO$_2$. The free OH can similarly recombine with hydrogen to reproduce H$_2$O, maintaining its high abundance in the CPD. CO$_2$ maintains an abundance ratio relative to H$_2$O of about 4\% which is lower than is found in the ice in the PPD. Earlier on in its evolution the disk streamline passes through a warm-spot in the PPD between 0-10 orbits that briefly sublimates the CO$_2$. This warm-spot can be seen in figure \ref{fig:temp} and corresponds to the streamline passing across the spiral wave of the embedded planet.

The bottom right panel shows the most abundant `organic' species that was formed in the CPD, formic acid. \cite{Cleeves15} model the chemical impact of an accreting proto-planet on the surrounding PPD by computing excess heating on the PPD gas due to the planet's accretion luminosity. They find that this excess radiation locally heats the PPD and produces an azimuthally asymmetric enhancement in H$_2$CO emission. Instead, in our work we achieve a higher temperatures (few 100s K) deep in the gravitational potential of the embedded planet compared to \cite{Cleeves15} (60-120 K) which drives our chemical evolution to a more oxidized organic species.

\begin{figure*}
    \centering
    \includegraphics[width=0.8\textwidth]{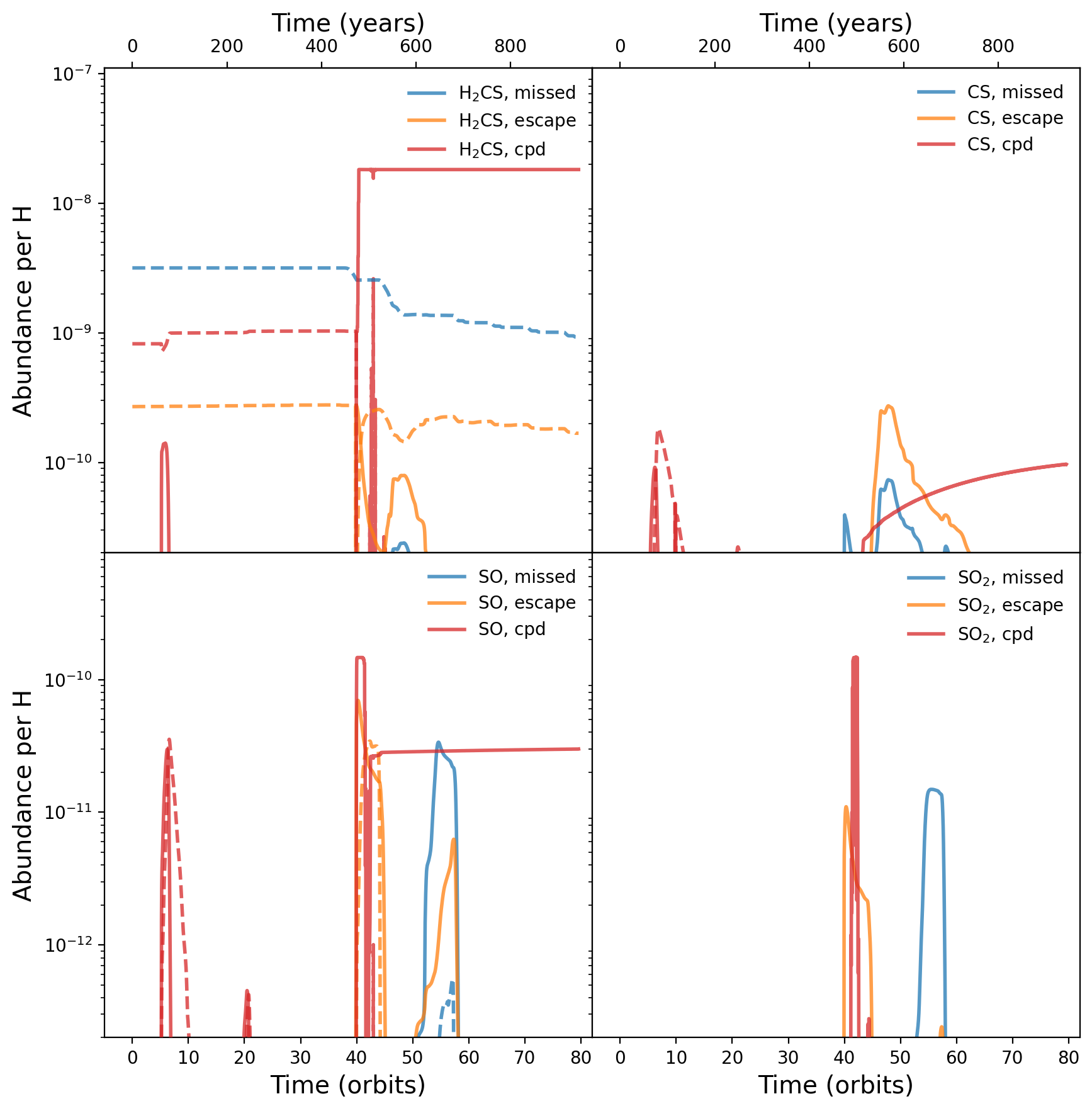}
    \caption{Same as in figure \ref{fig:Obearing} but for relevant sulfur bearing species. Note here that the axes range on the y-axis has been reduced by about 3 orders of magnitude compared to figure \ref{fig:Obearing}, and differ between the top and bottom rows.}
    \label{fig:Sbearing}
\end{figure*}

\subsubsection{Streamline chemistry: C-bearing species}

In figure \ref{fig:Cbearing} we show the dominant carbon-carrying species in the gas. Like in figure \ref{fig:Obearing} the left two panels are CO and CO$_2$, while the other two panels show methane (bottom right) and a larger carbon-chain molecule (top right). In the PPD ($t<40$ orbits) we find that methane remains in the gas phase, meaning that the gas is sufficiently warm to exceed the sublimation temperature of methane. This result differs from the physical setup of \cite{Jiang2023}, who's numerical simulations is setup with an embedded planet orbiting father from its host star than we do here. As such their CH$_4$ begins in the ice phase, and its sublimation is the key chemical process that drives important carbon chemistry in their work.

Here we see that carbon chemistry has already produced longer chain hydrocarbons (like C$_3$H$_4$, top right panel) in the PPD. These species begin in the ice phase, but are sublimated into the gas phase as the disk streamline enters into the gravitational influence of the embedded planet. Similar to H$_2$O, these long chain hydrocarbons have sublimation temperatures that are within the temperature fluctuation range of the disk streamlines. They thus continue to oscillate between gas and ice phases as the temperature changes. 

Methane spends most of its time as a gas and is greatly depleted by the presence of the CPD. In fact most of the methane is produced and maintained in streamlines that spend the most time away from the CPD. CH$_4$ has been observed in protoplanetary disks in the infrared \citep[for ex.][using NIRSPEC on Keck]{Gibb2013}, and recent observations in disks with JWST further suggest the presence of CH$_4$ gas emission in one system \citep{Tabone2023}.

\subsubsection{Streamline chemistry: S-bearing species}

Figure \ref{fig:Sbearing} shows the evolution of sulfur bearing species in the three different families. In these panels, the extent of the y-axes are four orders of magnitudes smaller than the axes of figures \ref{fig:Obearing} and \ref{fig:Cbearing} because sulfur is a less abundant element compared to both carbon and oxygen. While not shown in this figure, recall that in the PPD the sulfur is primarily in the form of H$_2$S ice (figure \ref{fig:initabun}). The top two panels show carbon rich sulfur-bearing species while the bottom two panels show oxygen rich ones. For the carbon rich species we see that the PPD starts with a small quantity of H$_2$CS in the ice phase which sublimates once the disk streamlines enters into the CPD (at t$>$40 orbits). Similarly, all three streamlines show an increase of CS during the closest approach to the planet, but it is sustained for the disk streamline and lost over $\sim 20$ orbits for the two streamlines that stay in the PPD.

Similar to the case of CO$_2$, when the disk streamline passes through the warm-spot H$_2$CS, CS, and SO are all produced in the gas phase during the briefly increase in temperature. When the streamline moves out of the warm spot the gas cools and all three species freeze out. For the oxygen rich sulfur-bearing species we find that their production and survival depends sensitively on the temperature of the gas. This includes temporary increases in temperature as was mentioned above, but also when the streamlines approach the embedded planet. The escaping streamline (orange) briefly produces both SO and SO$_2$ in the gas phase during its close approach to the planet, but they freeze out and destroyed when the gas cools. The disk streamline also produces both SO and SO$_2$ as it enters into the gravitational potential of the embedded planet. SO$_2$ is only maintained for a short time nearest to the embedded planet where the gas temperatures are the highests. At lower temperatures SO$_2$ can not be maintained and is rapidly depleted. SO, on the other hand is maintained in the CPD over the remainder of the integration time in the CPD. Later in the integration, both the missed and escaped streamline again pass through the spiral wake and breifly for SO and SO$_2$, but again it can not be maintained as the gas temperature returns to its colder temperature.

In the missed and escaped streamlines, the primary sulfur-bearing molecular species ends up being H$_2$S ice - just as it was after the original 1 Myr of chemical evolution in the PPD. The streamline that stays in the CPD, on the other hand, converts nearly all of its H$_2$S into SO, CS, and H$_2$CS. A small quantity of SO$_2$ is also produced but does not remain abundant in the colder regions of the CPD - it only exists in the CPD when gas temperatures exceed $\sim 500$K.  Like SO and CS, H$_2$CS has been detected in protoplanetary disks that feature a large dust cavity indicative of an embedded planet \citep[see for ex. ][]{Booth2023a}.

\subsection{Streamline chemistry: Nominal model}\label{sec:nominal}

The Nominal model from \cite{Legaetal} produces a warm gaseous, spherical envelope around the embedded planet rather than a CPD. Similar to the LowMass model, the streamline families can be split between streamlines that spend nearly their whole evolution inside the gravitational influence of the embedded planet, and streamlines that completely miss the gaseous envelope. We select a streamline that spends nearly all of its forward integration time in the gaseous envelope of the embedded planet (see table \ref{tab:initcon}) and analyze it in a similar manner as was done for the LowMass streamlines. 

In figure \ref{fig:nominaltemp} we compare the evolution of the gas temperature in the streamline of the nominal model to the disk streamline. Similar to the disk streamline, the streamline in the nominal model begins its forward integration at 40 orbits and slowly spirals closer to the embedded planet, to higher temperatures. In the backward integration (at $t<40$) the gas orbits in the corotating region of the embedded planet, and its temperature fluctuates between nearly 200 K and $\sim 20$ K. Once it enters into the envelope and spirals close to the embedded planet, its gas temperature fluctuates between 600-1000~K.

In figure \ref{fig:nominal} we show the evolution of H$_2$O, CO$_2$, SO, and SO$_2$ in the nominal streamline and compare that evolution to the disk streamline. At early times these molecules vary drastically as the gas temperature oscillates in the PPD at $t<40$ orbits. This also results in some production of SO in the PPD which oscillates along with the oscillating temperature in the PPD. As the streamline enters into the gaseous envelope surrounding the embedded planet it experiences the high temperatures shown in figure \ref{fig:nominaltemp}. SO is further enhanced in this high temperature gas as oxygen carriers like CO$_2$ are destroyed. The gas stays at these high temperatures so that the SO is eventually destroyed in favour of the more oxygen rich SO$_2$. A small fraction of the freed oxygen reacts with SO to produce SO$_2$ while the rest recombines with available hydrogen to produce H$_2$O. A warm, spherical gas envelope surrounding the embedded planet offers a unique chemical environment compare to the CPD, which generates much less SO$_2$, and only in the inner regions of the CPD. This difference in chemical behaviour and distribution of sulfur oxides may offer an observational means of differentiating between the difference gas accretion geometries. This would be worth exploring further in the future as it would allow for a more accessible observational means of studying the way that giant planets accrete their gas later in their formation.

\begin{figure}
    \centering
    \includegraphics[width=\linewidth]{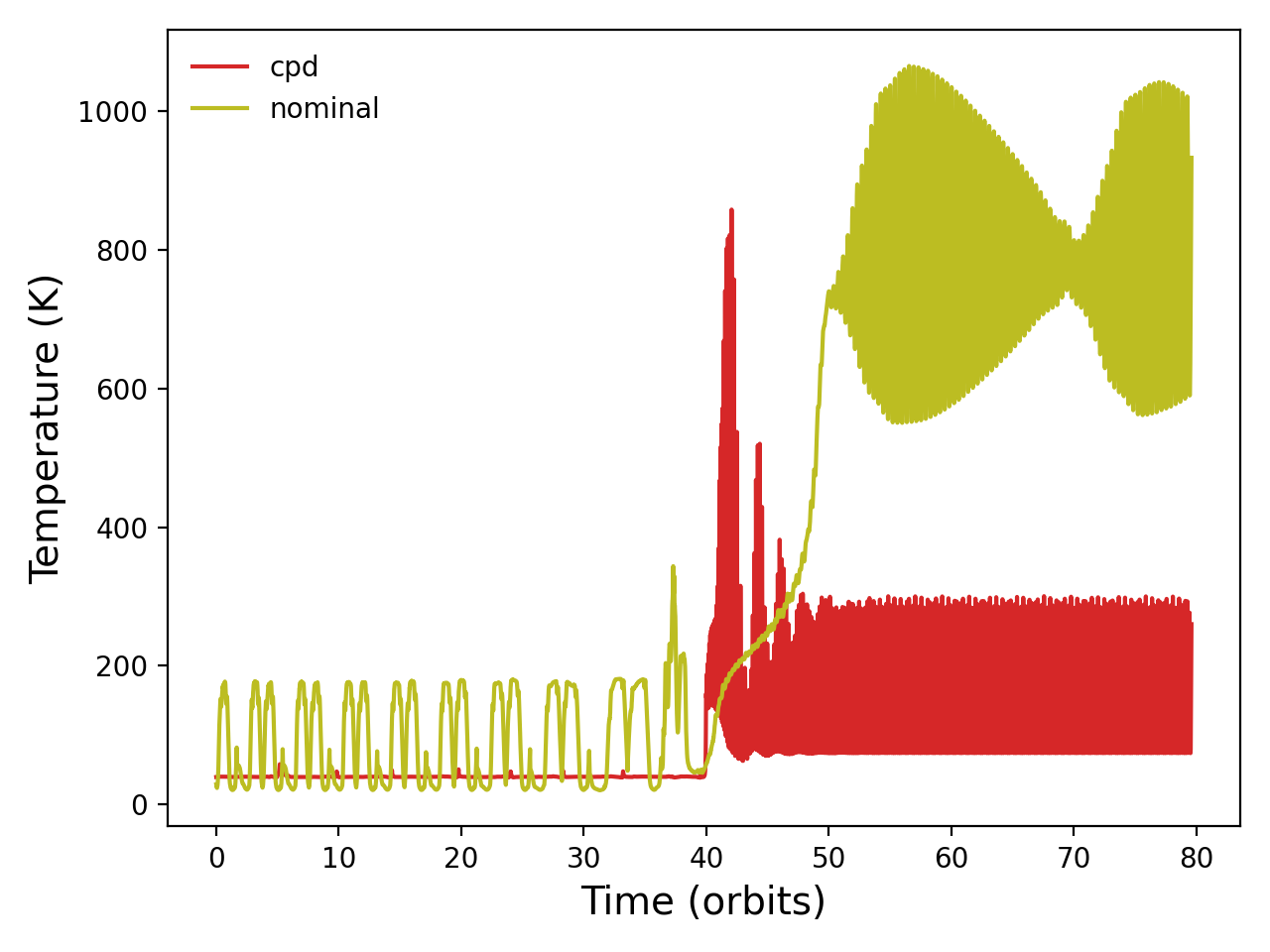}
    \caption{The evolution of the gas temperature of the streamline in the nominal model compared to the temperature evolution of the disk streamline.}
    \label{fig:nominaltemp}
\end{figure}

\begin{figure*}
    \centering
    \includegraphics[width=0.8\textwidth]{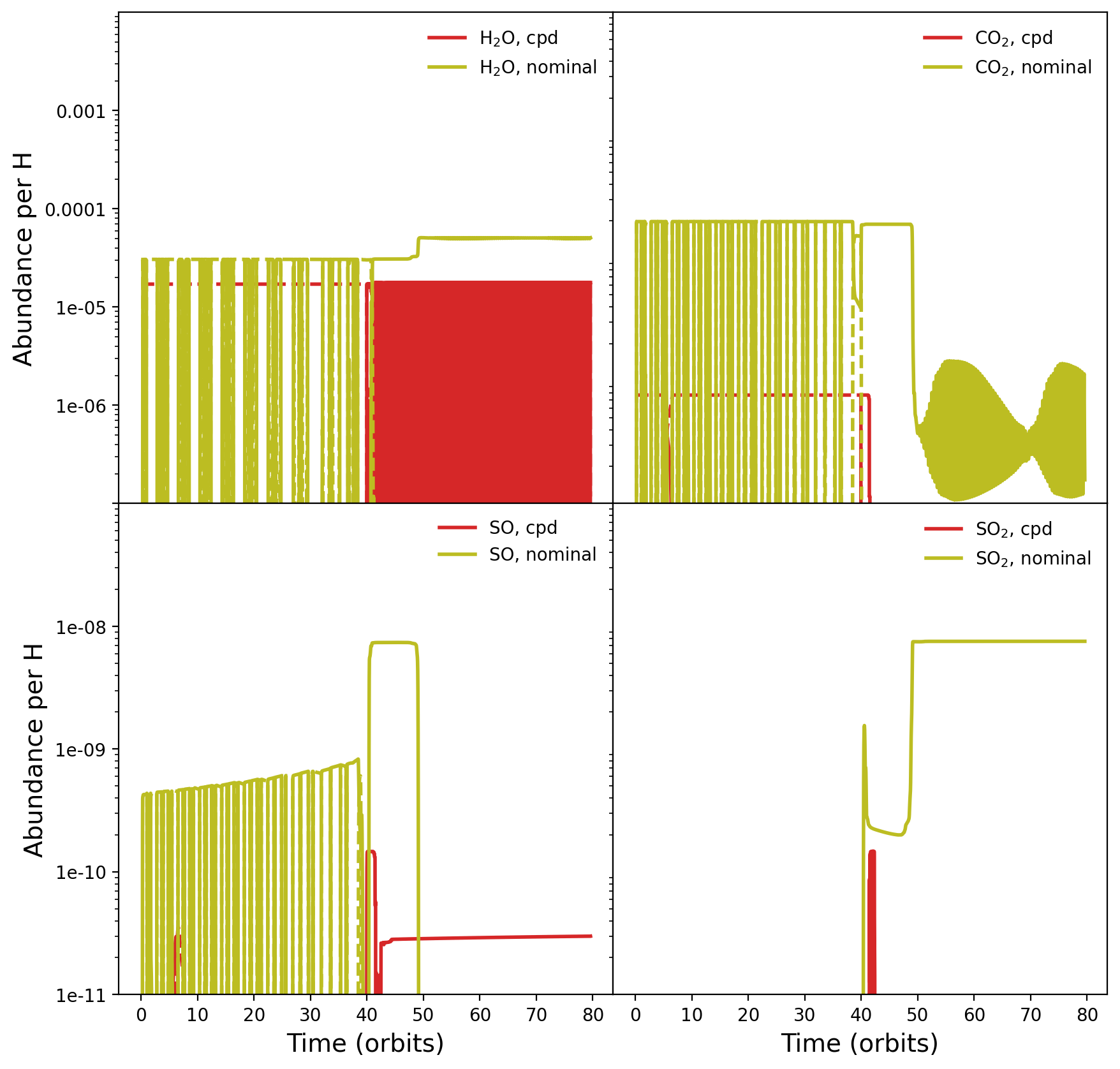}
    \caption{A comparison between a streamline in the nominal model and the LowMass model, using the disk streamline. As before the solid lines show the gas phase while the dashed lines show the ice phase.}
    \label{fig:nominal}
\end{figure*}

\section{Discussion}\label{sec:discussion}

\subsection{Chemical evolution of accreting gas}

\cite{Cridland2020} studied the impact on giant planet atmospheric chemistry by the accretion of gas via meridional flow. They sampled the chemical abundances of abundant volatile species (in both the gas and ice phase) in the incoming material at a range of heights above the disk midplane at the gap edge, and assumed that there was no chemical evolution between the gap edge and the accretion of the gas onto the planet. Here we have extended their work and computed the chemical evolution of gas along streamlines from the PPD into the CPD.

Due to the large temperature gradient between the PPD and the CPD all of the frozen volatile species are released into the gas phase very rapidly. This drives the gas to be more oxygen rich (relative to carbon) than it was in the PPD. In addition, since the gas temperature exceeds, for a short time, the sublimation temperature of refractory carbon \citep[$\sim$500 K;][]{Berg15} it is also possible that some of this extra carbon source could be released into the gas. The effect of refractory carbon on the final atmospheric chemistry of giant planet atmosphere has been explored by \cite{Crid19b}, however for simplicity we have ignored refractory carbon here. It is possible that the extra carbon released from this refractory source could act as a sync for the freed OH what otherwise reacted to produce SO and SO$_2$, possibly reducing their abundance close to the young planet.

Both here and in \cite{Cridland2020} it was assumed that the small (well coupled) dust grains followed the same dynamical evolution as the gas. In \cite{Cridland2020} this implied that volatile species like H$_2$O could accrete into a giant planet's atmosphere as ice species frozen on micron-sized dust grains. However, we have seen here that a more accurate expectation is that frozen-out volatiles should be released into the gas as it travels from the PPD, furthermore we might expect that the dust grains will grow while in the CPD \citep[as explored by][among others]{Drazkowska2018,BatyginMorbi2020,Shibaike2023}, and become decoupled to the gas. This dust growth could, in the future, drive moon formation, but will also lead to less efficient volatile freeze-out as the available surface area per mass of the dust is reduced.

With that said, it is clear from this work that in considering the delivery of important elemental species like C, O, N, and S to the atmosphere of the growing planet, we may assume that all volatile species (whether they are in the gas or ice phase) will follow the flow of the gas after entering into the CPD. This conclusion requires the gas to reach high enough temperatures ($\gtrsim 100$K) to sublimate all of the volatiles, which was easily reached here but could be more difficult as the mass accretion rate onto the CPD drops and the disk around the planet cools. This could thus imply that older planet-forming systems will generate fewer oxides because their water will remain in the ice phases. 

\subsection{Formation and structure of the CPD}

\begin{figure}
    \centering
     \includegraphics[width=0.5\textwidth]{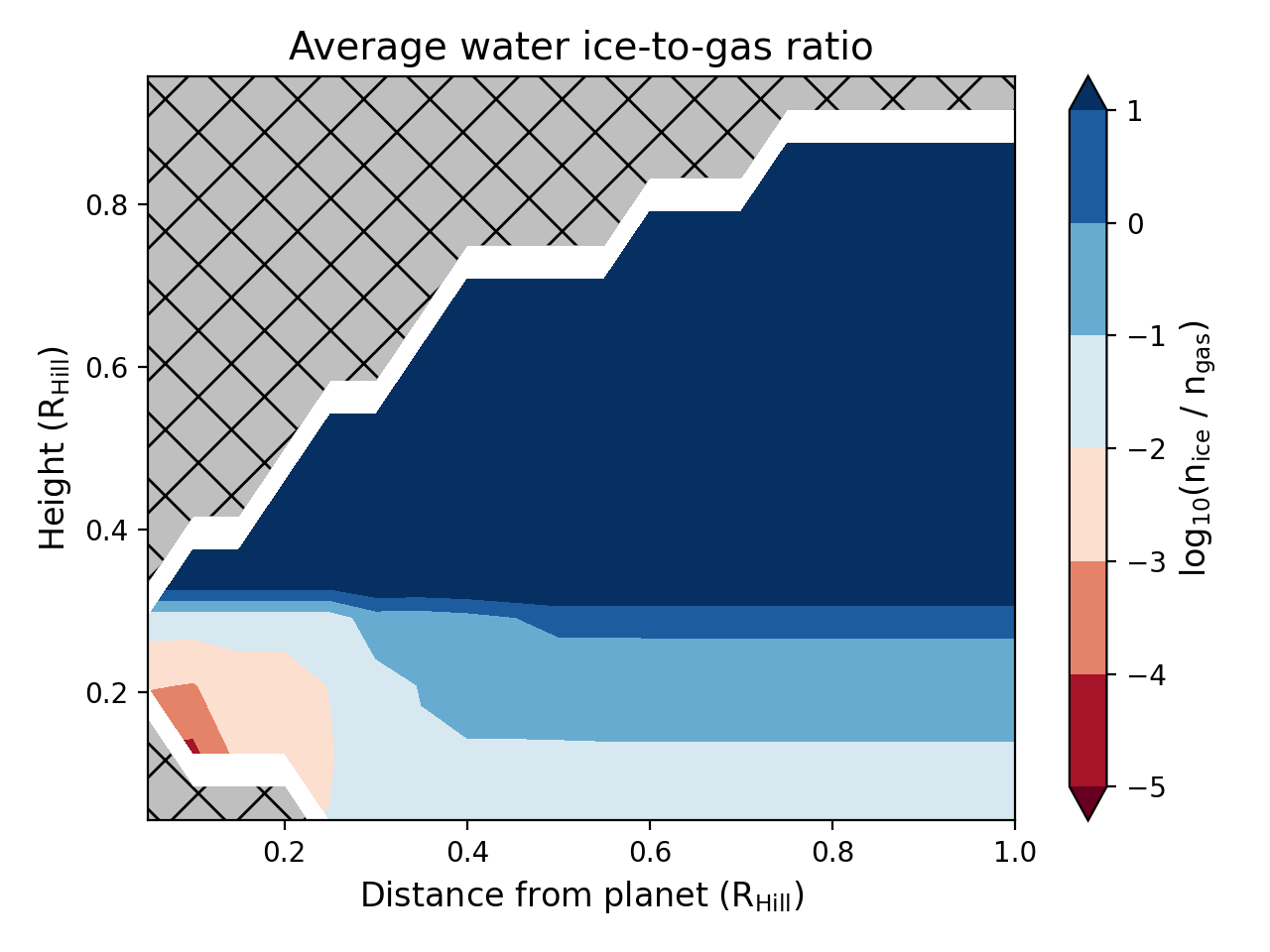}
    \caption{The aziumthally averaged H$_2$O ice-to-gas mass ratio of the CPD and material surrounding the embedded planet. Recall that the CPD extends to $\sim 0.4$ R$_{\rm Hill}$. We base the calculation on the disk streamline and thus the hatched region shows where the streamline did not visit during the simulation. The traditional ice line, where the abundance of gas and ice species are equal exists above the disk midplane, suggesting that no ice lines exist in the CPD.}
    \label{fig:water-solid-to-gas}   
    \includegraphics[width=0.5\textwidth]{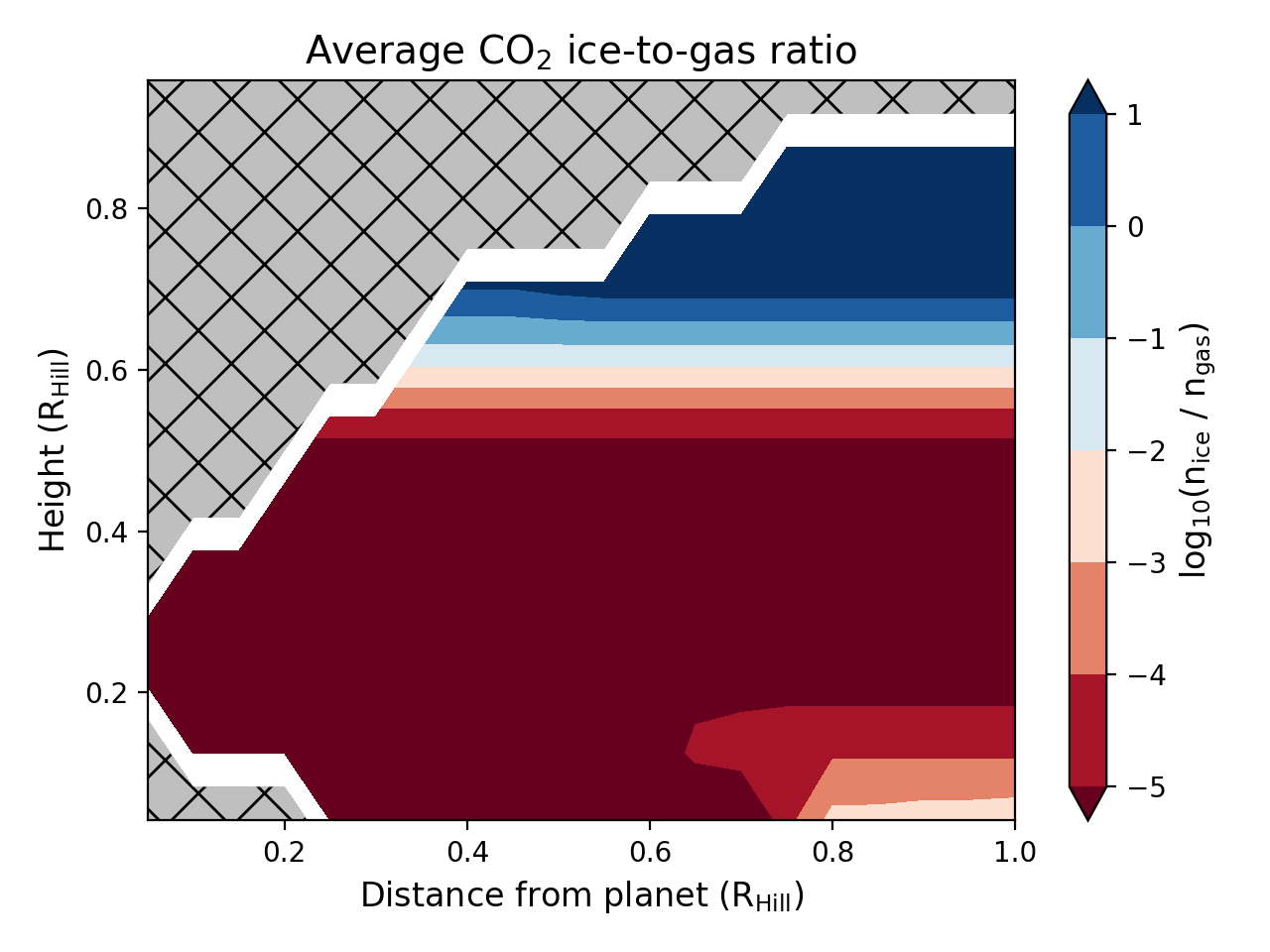}
    \caption{The aziumthally averaged CO$_2$ ice-to-gas mass ratio of the CPD and material surrounding the embedded planet. The traditional ice line, where the abundance of gas and ice species are equal lies well above the disk midplane due to the warm nature of the circumplanetary material.}
    \label{fig:co2-solid-to-gas}
\end{figure}

Our physical model represents a protoplanetary disk that is depleted in gas density compared to the MMSN by a factor of 10. As previously mentioned this can correspond to a more evolved disk that has accreted a significant amount of its mass either into the host star or into the formation of giant planets. As found in Paper 1, the LowMass model produces a CPD because the gas flux into the planet's Hill sphere was sufficiently low that it could cool in time to collapse into a rotationally support structure. In the Nominal model the protoplanetary disk was initialized with the standard MMSN, but rather than produce a rotationally supported disk it instead produces a pressure supported spherical envelope around the embedded planet. This result corresponds to the ones found in \citep{KlahrKley2006,Szulagyi2016,Fung2019,LLNCM19,2024arXiv240214638K} where the gas equation of state was treated in a similar manner to what was done in Paper 1.

In figure \ref{fig:water-solid-to-gas} we show a 2D view of the azimuthally averaged mass ratio of the water ice and gas species in the CPD, based on the disk streamline.  The grey hatches represent the space around the embedded planet where that streamline does not visit during the simulation. Given the warm nature of the CPD, we find that the gas phase dominates the abundance of volatile species, including when we analyze individual molecules. In our simulated CPD we lack any so-called `ice line', along the midplane of the disk because its temperature is too warm for volatile freeze out there. Higher above the CPD, in a shallower part of the gravitational potential the gas is cooler and water can thus freeze out. Likewise we find that the CO$_2$ ice line exists at an even higher height above the midplane, in a further colder area around the embedded planet.

This could complicate the view that ice lines, particularly the water ice line, play an important role in the formation of the Jovian moons \citep[see for ex.][]{Heller_2015}. Alternatively this suggests that moon formation from CPDs could occur later in the planet formation process, at a stage when the CPD is effectively cut off from future gas flow from the PPD. At this later stage, the CPD could cool which would allow for the emergence of the water ice line along the midplane of the CPD. A late stage of moon formation was argued by \cite{CW2002}, who found that Galilean moon formation was most consistent with mass accretion rates from the PPD to the CPD of $2\times 10^{-7}$ Jupiter masses per year. Over a sphere of radius 1 R$_{\rm Hill}$ we find that the mass inflow rate in the LowMass model is $1.8\times 10^{-5}$ Jupiter masses per year, two orders of magnitude higher than predicted by \cite{CW2002}. Our simulated CPD thus represents an early state of a CPD, where it is too warm and has too large of accretion rate to likely lead to moon formation. 

We note here that our model of focus - the LowMass model - represents a state where the gas density of the PPD is already reduced by an order of magnitude relative to the Nominal model (recall table \ref{tab:Simparams} and Paper 1). As previously implied, the LowMass model could represent an older stage of PPD evolution than the PPD initialized in the Nominal model. However as previously noted our modeled mass accretion rate into the region of the planet's gravitational influence is two orders of magnitude higher than was predicted in moon formation studies, which further emphasises the delayed nature of moon formation. This may imply that young systems that have bright, localized SO emission, like HD 169142 \citep{Law2023}, do not have ongoing moon formation around their embedded planet, while older systems like PDS 70 that has very low ongoing mass accretion \citep[as suggested by][]{Haffert2019} could be actively producing moons around their giant planets.

\subsection{Observational tracers of CPDs}

Given the recent interest in detecting planets in younger stages of evolution - the so-called proto-planet phase - as well as their presumed circumplanetary material, we explore what chemical tracers may be relevant to detecting CPDs. From figures \ref{fig:Obearing}-\ref{fig:Sbearing} we saw that many volatile species are in their gas phase in the CPD thanks to the much higher gas temperatures in the CPD (at least $\le 200$ K) compared to the surrounding PPD. The most abundant molecular species - CO - shows very little variation between being inside and outside of the CPD, and would thus be a poor tracer. This is due to the higher temperature of the protoplanetary disk gas that surrounds the embedded planet, which exceeds the freeze-out temperature of CO.

Other abundant species like H$_2$O and CO$_2$ are in the ice phase in the PPD - the embedded planet orbits sufficiently far enough away that the gas is colder than their freeze-out temperature. Meanwhile in the CPD the gas temperature exceeds their freeze-out temperature, and they exist mainly in the gas phase - indeed CO$_2$ is first destroyed at the highest CPD temperatures but is then produced in the CPD through the oxidation of CO by OH. H$_2$O is thermally dissociated, producing the needed OH for the aforementioned oxidation reaction, and elemental hydrogen. Because of the high abundance of hydrogen, many of the freed OH react to reproduce H$_2$O. The chemistry in the CPD is not so different from the chemistry in the inner few AU of the PPD and as such the inner disk may turn out to outshine the CPD.  

Sulfur-bearing species offer a potential way of differentiating between the CPD and PPD. SO and SO$_2$ are known tracers of high temperature gas - mainly in astrophysical shocks. Here (recall fig. \ref{fig:Sbearing}), we see that the gas becomes more abundant in SO when the streamlines orbit in the CPD, and these species are effectively obsolete when the gas is in the colder PPD midplane. A very common molecular carrier of sulfur, H$_2$S remains primarily in the ice phase throughout the evolution of the streamlines, and is rapidly destroyed when the gas temperature exceeds $\sim 200$ K. In the CPD the conversion from H$_2$S ice to species like CS, SO, and SO$_2$ is driven by high temperature chemistry, and the availability of oxidizers such as OH from the desorption and destruction of H$_2$O ice. There have been a small collection of observations of these heavier sulfur species made in protoplanetary disks. Chemical models have shown that the ratio of CS/SO is very sensitive to the local gas C/O \citep{Semenov2018}, and they have thus been primary targets to chemically characterise protoplanetary disks \citep{Booth2021,Keyte2023,Law2023,Booth2023a,Booth2024}. Interestingly, all of these observations show azimuthal asymmetries in the emission of sulfur-bearing species and many of these asymmetries have been attributed to the presence of young, embedded, planets. In particularly HD 169142, an SO emission `blob' has been found co-located with the proposed location of a young planet that has been inferred from other observational methods \citep{Law2023}. The possible link between oxidized sulfur species and the presence of embedded proto-planets is an interesting line of research for which we find some evidence here.

\subsection{Finding the needle in the haystack}

From their high abundance and relative uniqueness, we might expect that CPDs shine brighter in SO and SO$_2$ emission lines than the surrounding gas in the PPD. This is complicated however, by the fact that the warm gas in the upper layers of the PPD will also produce some amount of oxidized sulfur species. In face-on disks, the CPD may represent an asymmetry in the emission structure of SO and SO$_2$ which could help to localize the position of young proto-planets and their circumplanetary material. 

At the end of section \ref{sec:methods} we outline our method of estimating the column number density of the CPD from the disk streamline. The column densities implied by our streamline calculation on a length scale of 1 Hill radius provides estimates of $2.0\times 10^{14}$, $7.3\times 10^{14}$, $4.6\times 10^{14}$, and $1.7\times 10^{17}$ cm$^{-2}$ in the CPD for SO, SO$_2$, CS, and H$_2$CS respectively. These represent the total column number densities from the midplane of the CPD up towards the observer assuming that the system is observed face-on. Furthermore this calculation ignores any optical depth effects due to the presence of dust grains. These calculation thus represent the maximum possible column number density that could be inferred observationally.

As a first step in understanding the detectability of CPDs in young stellar systems we turn to the well known PDS 70 system \citep{Keppler2018,Benisty2021}. Recent sub-mm emission has suggested that PDS 70c has some surrounding circum-planetary material while PDS 70b does not \citep{Benisty2021}. \cite{Crid2023} explored a chemical model of the PPD around PDS 70 in order to understand the chemical properties of the gas that built the PDS 70b and PDS 70c planets. Their chemical model used the thermo-chemical code DALI \citep{Bruderer2012,Bruderer2013} and used the chemical network of \cite{Miotello2019} which focused on carbon chemistry and the production of small chain hydrocarbons, but lacked any sulfur chemistry.

To facilitate a comparison between the gas column densities inferred in the CPD of this model and the gas column densities inferred in the PPD surrounding PDS 70, we recompute the chemistry of the `best fit' physical model of \cite{Crid2023} using an updated chemical network. This network combines the network used in \cite{Crid2023} with a set of 330 sulfur-relevant reactions that are included in the \verb|osu_03_2008| network. In this way the carbon chemistry, which was the main driver for finding the best fit physical model in \cite{Crid2023}, should remain largely unchanged while the sulfur chemistry of the PDS 70 disk can be newly analysed.

\begin{figure*}
    \centering
    \includegraphics[width=\textwidth]{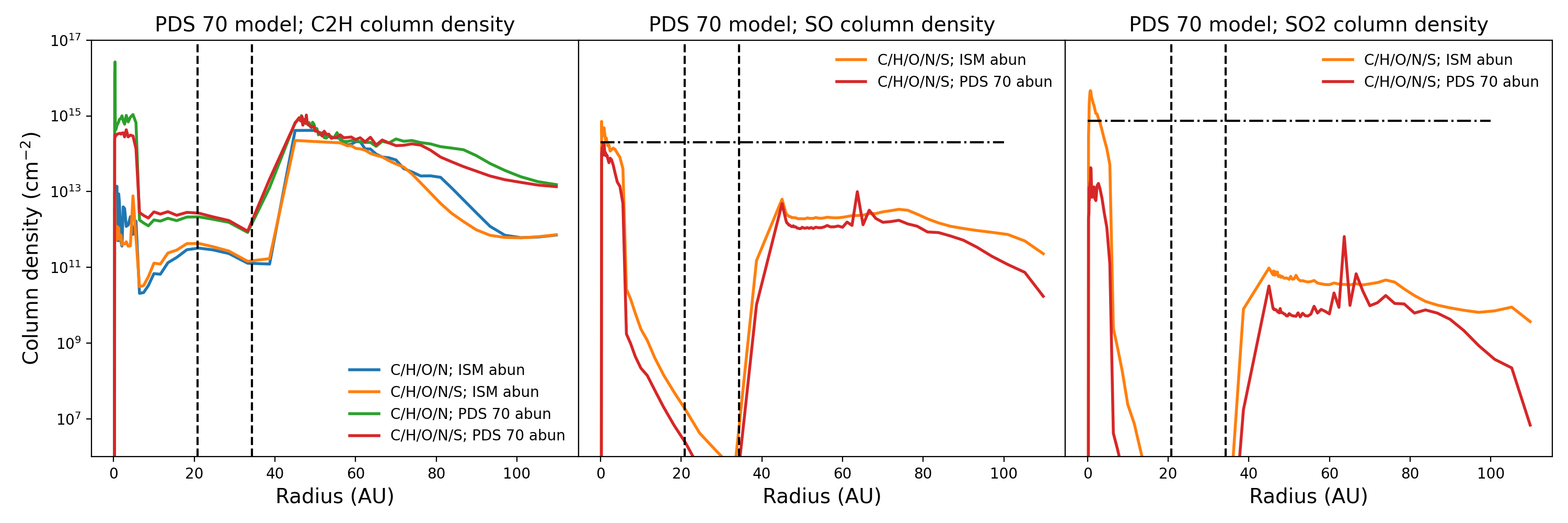}
    \caption{\textbf{Left}: A comparison between the C$_2$H column density between the different chemical networks mentioned in the text. The networks are \cite{Miotello2019}: C/H/O/N and \cite{Miotello2019} $+$ \cite{Bruderer2013}: C/H/O/N/S. We tested that the resulting column density showed minimal change when adding the extra sulfur reactions to the network of \cite{Miotello2019}. There is minimal change in the column density for the different networks regardless of the elemental abundance of carbon and oxygen, which we vary between the ISM value (C/O $=0.4$) and the abundance inferred for PDS 70 (C/O $=1.01$) by \cite{Crid2023}. The dashed lines show the orbital positions of the b and c planets. \textbf{Middle}: Same as \textbf{Left} but for SO. Note that in the \cite{Miotello2019} network SO and other more complex sulfur-bearing species are not included. The models shown in \textbf{Left} corresponding to that network are thus left out here. The dot-dashed line shows the estimated column density from our chemical model along the streamlines. \textbf{Right}: Same as \textbf{Middle} but for SO$_2$. }
    \label{fig:C2Hcol}
\end{figure*}

We first check that adding the extra reactions to the network used in \cite{Crid2023} does not greatly impact the results of that work (see the left panel of Figure \ref{fig:C2Hcol}). In the figure we show a comparison between the two networks (labelled by the elements they include) for different assumed elemental abundances of carbon and oxygen, the sulfur abundance is kept constant at S/H = $1.91\times 10^{-8}$ throughout. In the left panel of Figure \ref{fig:C2Hcol} we can see that the addition of the sulfur-relevant reactions has not greatly changed the distribution of the C$_2$H column that was primarily used to restrict the physical model of the PDS 70 disk. 

An important feature of the PDS 70 disk is that it has a deep gap in the dust and gas which has been carved out by the young PDS\,70b and c planets. This gap can be seen in figure \ref{fig:C2Hcol} as the drop of about 3-4 orders of magnitude in the C$_2$H column density. Such a deep gap may further simplify the detection of a CPD in molecular emission because the surrounding PPD emission is suppressed by the low column density.

In the middle and right panels of Figure \ref{fig:C2Hcol} we show the column densities of the SO and SO$_2$ in the full network respectively. We see a similar drop in the column density at the location of the combined planetary gap. In both cases we see an even larger drop in the column density relative to the abundances in the inner and outer disk than was seen in C$_2$H. This could imply an even greater contrast between the PPD and CPD SO$_2$ emission. The column number densities in the CPD are about a factor of 10 larger than the SO and SO$_2$ column of the gas in the outer disk inferred in the PDS 70 models (see the dot-dashed lines in Figure \ref{fig:C2Hcol}). This implies that their emission should be brighter compared to the rest of the PPD.

Recently \cite{Rampinelli2024} report a weak detection of SO emission from the PDS 70 disk in a spacial resolution chemical survey of the disk. The peak emission, however, does not coincide with the positions either PDS 70b nor PDS 70c and instead peaks farther from the host star than the continuum emission ring. The emission non-detection near the PDS 70 proto-planets implies that any circum-planetary material that may be around them is not sufficiently hot to produce SO chemical, nor excite the specific lines observed. Given that PDS 70 is an older system this result is not surprising because the ongoing planetary accretion rate \citep[as inferred from H-alpha emission][]{Haffert2019} is very low.

For slightly younger systems the picture is more complicated. Recently \cite{Booth2023a} reported the detection of SO emission from the planet-hosting Herbig star HD 100546, while \cite{Booth2023b} report SO emission from the planet-hosting young system HD 169142. In both cases the emission is asymmetric around the disk's rotation axis and shows signs that the asymmetries are related to the presence of the embedded planets. The column number density of SO is estimated to be $2.1\times 10^{13}$ cm$^{-2}$ over the whole disk in HD 169142 and $4.9\times 10^{13}$ cm$^{-2}$ in the north-east `blob' \citep{Law2023}. This estimate in the `blob', at the position of a suspected planet, is a factor of 4 lower than our estimate from the streamline chemistry. This discrepancy can be easily explained by the presence of dust providing a source of opacity to shield some of the emission from these molecular species.

Nevertheless we expect that the emission of some of these sulfur-bearing species in young systems containing an embedded planet and circumplanetary material will be asymmetric around their rotation axis. This could lead to a separate test of the presence of young protoplanets and CPDs to reinforce other methods.

\subsection{Caveats and future prospects}

As a first step to understanding the chemical properties of the gas flowing from the PPD to the CPD we have taken a single snapshot of the velocity field to compute our streamlines. We have checked in Paper 1 that the simulation had reached a steady state, however in principle the velocity field may change slightly over a dynamical time. A given streamline may thus shift slightly from its computed trajectory shown here. While this might impact an individual streamline, it does not greatly impact the properties of the gas that control the chemistry - mainly the gas temperature and density - and so shouldn't change our overall conclusions.

\subsubsection{Diffusion}

We treat each streamline as an individual packet of gas with small (well coupled) dust grains. We ignore the impact of diffusion between a gas packet and the surrounding gas, which may have an impact on the evolution of the gas as it moves from the PPD to the CPD. In particular this effect may be greatest in the CPD where the temperatures / densities are the highest. Diffusion may mix volatiles like H$_2$O out of a gas packet, decaying the desorption / absorption oscillation that we see in figure \ref{fig:Obearing}. Removing H$_2$O from the gas packet reduces the production of OH and subsequently may lead to less efficient production of SO in the CPD, leading to a slight reduction in its line emission brightness.

\subsubsection{Heating and cooling}

We ignore the radiative heating of the forming planet on the gas temperature and high energy photon flux (impacting photodissociation and ionization). The young planet radiatively heats both the gas and dust in the CPD as well as the gas in the PPD nearby. This has been explored by many past authors \citep[for ex.][explored the impact of accretion heating on the water ice line location in the CPD]{Heller_2015}, in particular the work of \cite{Cleeves15} who suggest that azimuthal asymmetries in the molecular line emission of some voaltiles should arise from the local accretional heating of the growing planet. Further more, the thermal effects within the Hill sphere and the thermal structure of the disk have been explored by \cite{Jiang2023} and \cite{Chen2024} respectively. These effects could change the overall chemical evolution of the gas as it flows, and we leave this to future work.

Along with the radiative flux, the gas can be shock heated as it compacts the accretion planet, its magnetic field, and/or the CPD. We include (see Paper 1) compressional heating, which follows $P\nabla\cdot\textbf{v}$, due to the converging gas flows. However with finite spacial resolution it is possible that we do not fully resolve the shock front and miss the maximum temperature within the shock. In the shock chemistry model of \cite{vanGelder2021} SO also tends to be more abundant than SO$_2$ at the end of the shock which is similar to our results here. Future work could explore the energetics of the line emission which could depend on the exact temperature structure of the shocking gas.

One common feature of the CPD streamlines is their oscillating temperature as they orbit the embedded planet. It is possible that the cooling rates within the gas parcel are not sufficiently fast to account for the rapid temperature oscillations that are inferred in our method. Using the standard \cite{BellLin1994} opacity dependence on density and temperature, the functional form of the cooling rate from Paper 1 ($\rho\kappa_P a_r T^4$)\footnote{$a_r$ is the radiation constant and $\kappa_P$ is the Planck opacity.}, and the assumption that the gas energy density ($e$) follows $P = (\gamma-1)e$ for an ideal gas with $\gamma = 1.4$ we estimate a cooling timescale between 0.1 and 0.2 orbits. The gas streamlines orbit the embedded planet roughly 6 times per orbit of the planet, meaning that their orbital timescale is approximately 0.16 orbits. 

With similar timescales, it is unlikely that the gas would be able to cool from $\sim 300$ K down to $\sim 80$ K within a single orbit which would imply that the gas would stay nearer to the higher of the two temperatures during the orbit of the streamline. This may not change our ultimate conclusions because the important chemical processes occur when the temperature is at its highest, when the water is in its gas phase. 

\subsubsection{High energy photons}

Finally we note that the process of accretion is generally linked to an enhanced UV flux due to shocking gases and high temperatures. We have ignored any extra UV flux from the accreting planet in computing the local radiation field and its effect on the chemical evolution. We note that the temporal evolution of the accreting planet's luminosity has been modelled in the past \citep[for ex. by][]{Mordasini2013} and the accretion history of the planet could impact the ongoing chemical evolution of the gas in the CPD. A more thorough investigation over a longer timeframe is needed, including an investigation of the chemical properties of the circumplanetary material when the embedded planet is less massive and thus has a shallower gravitational potential.

\section{Conclusion}\label{sec:conclusion}

In this work we have post-processed numerical simulations of gas flow around an embedded Jupiter-massed planet. We compute the chemical evolution of the gas as it flows from the PPD into the gravitational influence of the embedded planet. For our selected simulation, the LowMass model of Paper 1, the circumplanetary material is organized into a CPD. There is a significant (nearly 800 K) temperature gradient between the PPD and the warmest regions of the CPD which releases all of the volatile ices from the coupled dust grains into the gas phase. The high temperatures and releases of frozen species result in high temperature chemistry which drives the formation of sulfur-bearing species that are otherwise absent in our simulated PPD. We generally find that:
\begin{itemize}
    \item Abundant volatiles like H$_2$O, CO$_2$, and H$_2$S are released into the gas phase, making the gas more oxygen and sulfur rich.
    \item CH$_4$ is dissociated in the warm gas around the embedded disk, allowing for the production of more long-chain hydrocarbons which exist in both the gas and ice phases in the CPD.
    \item Both carbon- and oxygen-bearing sulfur species are produced in the warm temperatures around the embedded planet, and are maintained in the gas phase.
    \item The high temperatures of the CPD exclude the existence of any canonical volatile ice lines along the midplane of the CPD. The lack of these lines limit the possibility of moon formation within our modelled CPD.
    \item The possible contrast in column number densities between the CPD and surrounding PPD, SO, SO$_2$, and CS could be used to trace the presence of young, embedded protoplanets in their natal PPD by looking for bright azimuthal asymmetries in the emission of those molecules.
    \item The systems that show strong emission in the aforementioned molecules are likely in an intermediate state of formation. Occurring at an age when the surrounding PPD has evolved sufficiently to lower its mass accretion rate by an order of magnitude, but early enough that the CPD has not sufficiently cooled to allow for the production of moons. 
\end{itemize}
Based on the empirical fit of \cite{Hartmann1998} an order of magnitude reduction of mass accretion rate relates to an age of a few Myr, while the only confirmed example of protoplanets with an accompanying natal disk (PDS 70) is at least 5 Myr old and has a remarkably low mass accretion \citep{Haffert2019}. Our model thus best represents protoplanet / protoplanetary disks systems that are on the order of a few Myrs old, at a stage when the circumplanetary material is still quite warm. 

\begin{acknowledgements}

  AC and MB acknowledges funding from the European Research Council (ERC) under the European Union’s Horizon 2020 research and innovation programme (PROTOPLANETS, grant agreement No. 101002188). Exoplanet studies at LMU are done as part of the Theoretical Astrophysics of Extrasolar Planets chair. This project made use of the following software: Astropy \citep{astropy}, SciPy \citep{Scipy}, NumPy \citep{numpy}, and Matplotlib \citep{matplotlib}. 

\end{acknowledgements}

\bibliographystyle{aa} 
\bibliography{mybib.bib} 

\begin{thebibliography}{81}
\expandafter\ifx\csname natexlab\endcsname\relax\def\natexlab#1{#1}\fi

\bibitem[{{Astropy Collaboration} {et~al.}(2022){Astropy Collaboration},
  {Price-Whelan}, {Lim}, {Earl}, {Starkman}, {Bradley}, {Shupe}, {Patil},
  {Corrales}, {Brasseur}, {N{\"o}the}, {Donath}, {Tollerud}, {Morris},
  {Ginsburg}, {Vaher}, {Weaver}, {Tocknell}, {Jamieson}, {van Kerkwijk},
  {Robitaille}, {Merry}, {Bachetti}, {G{\"u}nther}, {Aldcroft},
  {Alvarado-Montes}, {Archibald}, {B{\'o}di}, {Bapat}, {Barentsen},
  {Baz{\'a}n}, {Biswas}, {Boquien}, {Burke}, {Cara}, {Cara}, {Conroy},
  {Conseil}, {Craig}, {Cross}, {Cruz}, {D'Eugenio}, {Dencheva}, {Devillepoix},
  {Dietrich}, {Eigenbrot}, {Erben}, {Ferreira}, {Foreman-Mackey}, {Fox},
  {Freij}, {Garg}, {Geda}, {Glattly}, {Gondhalekar}, {Gordon}, {Grant},
  {Greenfield}, {Groener}, {Guest}, {Gurovich}, {Handberg}, {Hart},
  {Hatfield-Dodds}, {Homeier}, {Hosseinzadeh}, {Jenness}, {Jones}, {Joseph},
  {Kalmbach}, {Karamehmetoglu}, {Ka{\l}uszy{\'n}ski}, {Kelley}, {Kern},
  {Kerzendorf}, {Koch}, {Kulumani}, {Lee}, {Ly}, {Ma}, {MacBride}, {Maljaars},
  {Muna}, {Murphy}, {Norman}, {O'Steen}, {Oman}, {Pacifici}, {Pascual},
  {Pascual-Granado}, {Patil}, {Perren}, {Pickering}, {Rastogi}, {Roulston},
  {Ryan}, {Rykoff}, {Sabater}, {Sakurikar}, {Salgado}, {Sanghi}, {Saunders},
  {Savchenko}, {Schwardt}, {Seifert-Eckert}, {Shih}, {Jain}, {Shukla}, {Sick},
  {Simpson}, {Singanamalla}, {Singer}, {Singhal}, {Sinha}, {Sip{\H{o}}cz},
  {Spitler}, {Stansby}, {Streicher}, {{\v{S}}umak}, {Swinbank}, {Taranu},
  {Tewary}, {Tremblay}, {Val-Borro}, {Van Kooten}, {Vasovi{\'c}}, {Verma}, {de
  Miranda Cardoso}, {Williams}, {Wilson}, {Winkel}, {Wood-Vasey}, {Xue},
  {Yoachim}, {Zhang}, {Zonca}, \& {Astropy Project Contributors}}]{astropy}
{Astropy Collaboration}, {Price-Whelan}, A.~M., {Lim}, P.~L., {et~al.} 2022,
  \apj, 935, 167

\bibitem[{{Ayliffe} \& {Bate}(2009)}]{Ayliffe2009}
{Ayliffe}, B.~A. \& {Bate}, M.~R. 2009, \mnras, 397, 657

\bibitem[{{Batygin} \& {Morbidelli}(2020)}]{BatyginMorbi2020}
{Batygin}, K. \& {Morbidelli}, A. 2020, \apj, 894, 143

\bibitem[{{Bell} \& {Lin}(1994)}]{BellLin1994}
{Bell}, K.~R. \& {Lin}, D.~N.~C. 1994, \apj, 427, 987

\bibitem[{{Benisty} {et~al.}(2021){Benisty}, {Bae}, {Facchini}, {Keppler},
  {Teague}, {Isella}, {Kurtovic}, {P{\'e}rez}, {Sierra}, {Andrews},
  {Carpenter}, {Czekala}, {Dominik}, {Henning}, {Menard}, {Pinilla}, \&
  {Zurlo}}]{Benisty2021}
{Benisty}, M., {Bae}, J., {Facchini}, S., {et~al.} 2021, \apjl, 916, L2

\bibitem[{{Bergin} {et~al.}(2015){Bergin}, {Blake}, {Ciesla}, {Hirschmann}, \&
  {Li}}]{Berg15}
{Bergin}, E.~A., {Blake}, G.~A., {Ciesla}, F., {Hirschmann}, M.~M., \& {Li}, J.
  2015, Proceedings of the National Academy of Science, 112, 8965

\bibitem[{{Bergin} \& {Snell}(2002)}]{Bergin2002}
{Bergin}, E.~A. \& {Snell}, R.~L. 2002, \apjl, 581, L105

\bibitem[{{Birnstiel} {et~al.}(2010){Birnstiel}, {Dullemond}, \&
  {Brauer}}]{B10}
{Birnstiel}, T., {Dullemond}, C.~P., \& {Brauer}, F. 2010, \aap, 513, A79

\bibitem[{{Booth} {et~al.}(2023{\natexlab{a}}){Booth}, {Ilee}, {Walsh}, {Kama},
  {Keyte}, {van Dishoeck}, \& {Nomura}}]{Booth2023a}
{Booth}, A.~S., {Ilee}, J.~D., {Walsh}, C., {et~al.} 2023{\natexlab{a}}, \aap,
  669, A53

\bibitem[{{Booth} {et~al.}(2023{\natexlab{b}}){Booth}, {Law}, {Temmink},
  {Leemker}, \& {Macias}}]{Booth2023b}
{Booth}, A.~S., {Law}, C.~J., {Temmink}, M., {Leemker}, M., \& {Macias}, E.
  2023{\natexlab{b}}, arXiv e-prints, arXiv:2308.07910

\bibitem[{{Booth} {et~al.}(2024){Booth}, {Temmink}, {van Dishoeck}, {Evans},
  {Ilee}, {Kama}, {Keyte}, {Law}, {Leemker}, {van der Marel}, {Nomura},
  {Notsu}, {{\"O}berg}, \& {Walsh}}]{Booth2024}
{Booth}, A.~S., {Temmink}, M., {van Dishoeck}, E.~F., {et~al.} 2024, \aj, 167,
  165

\bibitem[{{Booth} {et~al.}(2021){Booth}, {van der Marel}, {Leemker}, {van
  Dishoeck}, \& {Ohashi}}]{Booth2021}
{Booth}, A.~S., {van der Marel}, N., {Leemker}, M., {van Dishoeck}, E.~F., \&
  {Ohashi}, S. 2021, \aap, 651, L6

\bibitem[{{Bosman} {et~al.}(2021){Bosman}, {Alarc{\'o}n}, {Bergin}, {Zhang},
  {van't Hoff}, {{\"O}berg}, {Guzm{\'a}n}, {Walsh}, {Aikawa}, {Andrews},
  {Bergner}, {Booth}, {Cataldi}, {Cleeves}, {Czekala}, {Furuya}, {Huang},
  {Ilee}, {Law}, {Le Gal}, {Liu}, {Long}, {Loomis}, {M{\'e}nard}, {Nomura},
  {Qi}, {Schwarz}, {Teague}, {Tsukagoshi}, {Yamato}, \&
  {Wilner}}]{Bosman2021MAPS}
{Bosman}, A.~D., {Alarc{\'o}n}, F., {Bergin}, E.~A., {et~al.} 2021, \apjs, 257,
  7

\bibitem[{{Bosman} {et~al.}(2018){Bosman}, {Walsh}, \& {van
  Dishoeck}}]{Bosman2018}
{Bosman}, A.~D., {Walsh}, C., \& {van Dishoeck}, E.~F. 2018, \aap, 618, A182

\bibitem[{{Bruderer}(2013)}]{Bruderer2013}
{Bruderer}, S. 2013, \aap, 559, A46

\bibitem[{{Bruderer} {et~al.}(2012){Bruderer}, {van Dishoeck}, {Doty}, \&
  {Herczeg}}]{Bruderer2012}
{Bruderer}, S., {van Dishoeck}, E.~F., {Doty}, S.~D., \& {Herczeg}, G.~J. 2012,
  \aap, 541, A91

\bibitem[{{Canup} \& {Ward}(2002)}]{CW2002}
{Canup}, R.~M. \& {Ward}, W.~R. 2002, \aj, 124, 3404

\bibitem[{{Canup} \& {Ward}(2006)}]{CanupWard2006}
{Canup}, R.~M. \& {Ward}, W.~R. 2006, \nat, 441, 834

\bibitem[{{Canup} \& {Ward}(2009)}]{CanupWard2009}
{Canup}, R.~M. \& {Ward}, W.~R. 2009, in Europa, ed. R.~T. {Pappalardo}, W.~B.
  {McKinnon}, \& K.~K. {Khurana}, 59

\bibitem[{{Chen} {et~al.}(2024){Chen}, {Kama}, {Pinilla}, \&
  {Keyte}}]{Chen2024}
{Chen}, K., {Kama}, M., {Pinilla}, P., \& {Keyte}, L. 2024, \mnras, 527, 2049

\bibitem[{{Choksi} \& {Chiang}(2024)}]{ChoksiChiang2024}
{Choksi}, N. \& {Chiang}, E. 2024, arXiv e-prints, arXiv:2403.10057

\bibitem[{{Christiaens} {et~al.}(2019){Christiaens}, {Cantalloube}, {Casassus},
  {Price}, {Absil}, {Pinte}, {Girard}, \& {Montesinos}}]{Christiaens2019}
{Christiaens}, V., {Cantalloube}, F., {Casassus}, S., {et~al.} 2019, \apjl,
  877, L33

\bibitem[{{Cleeves} {et~al.}(2015){Cleeves}, {Bergin}, {Qi}, {Adams}, \&
  {{\"O}berg}}]{Cleeves15}
{Cleeves}, L.~I., {Bergin}, E.~A., {Qi}, C., {Adams}, F.~C., \& {{\"O}berg},
  K.~I. 2015, \apj, 799, 204

\bibitem[{{Cridland} {et~al.}(2020){Cridland}, {Bosman}, \& {van
  Dishoeck}}]{Cridland2020}
{Cridland}, A.~J., {Bosman}, A.~D., \& {van Dishoeck}, E.~F. 2020, \aap, 635,
  A68

\bibitem[{{Cridland} {et~al.}(2019{\natexlab{a}}){Cridland}, {Eistrup}, \& {van
  Dishoeck}}]{Crid19b}
{Cridland}, A.~J., {Eistrup}, C., \& {van Dishoeck}, E.~F. 2019{\natexlab{a}},
  \aap, 627, A127

\bibitem[{{Cridland} {et~al.}(2023){Cridland}, {Facchini}, {van Dishoeck}, \&
  {Benisty}}]{Crid2023}
{Cridland}, A.~J., {Facchini}, S., {van Dishoeck}, E.~F., \& {Benisty}, M.
  2023, \aap, 674, A211

\bibitem[{{Cridland} {et~al.}(2019{\natexlab{b}}){Cridland}, {van Dishoeck},
  {Alessi}, \& {Pudritz}}]{Crid19c}
{Cridland}, A.~J., {van Dishoeck}, E.~F., {Alessi}, M., \& {Pudritz}, R.~E.
  2019{\natexlab{b}}, \aap, 632, A63

\bibitem[{{D'Alessio} {et~al.}(2006){D'Alessio}, {Calvet}, {Hartmann},
  {Franco-Hern{\'a}ndez}, \& {Serv{\'{\i}}n}}]{Dal2006}
{D'Alessio}, P., {Calvet}, N., {Hartmann}, L., {Franco-Hern{\'a}ndez}, R., \&
  {Serv{\'{\i}}n}, H. 2006, \apj, 638, 314

\bibitem[{{de Val-Borro} {et~al.}(2006){de Val-Borro}, {Edgar}, {Artymowicz},
  {Ciecielag}, {Cresswell}, {D'Angelo}, {Delgado-Donate}, {Dirksen}, {Fromang},
  {Gawryszczak}, {Klahr}, {Kley}, {Lyra}, {Masset}, {Mellema}, {Nelson},
  {Paardekooper}, {Peplinski}, {Pierens}, {Plewa}, {Rice}, {Sch{\"a}fer}, \&
  {Speith}}]{deValBorroetal2006}
{de Val-Borro}, M., {Edgar}, R.~G., {Artymowicz}, P., {et~al.} 2006, \mnras,
  370, 529

\bibitem[{{Draine}(1978)}]{Draine1978}
{Draine}, B.~T. 1978, \apjs, 36, 595

\bibitem[{{Dr{\k{a}}{\.z}kowska} \& {Dullemond}(2018)}]{Drazkowska2018}
{Dr{\k{a}}{\.z}kowska}, J. \& {Dullemond}, C.~P. 2018, \aap, 614, A62

\bibitem[{{Dullemond} \& {Dominik}(2005)}]{DD05}
{Dullemond}, C.~P. \& {Dominik}, C. 2005, \aap, 434, 971

\bibitem[{{Eistrup} {et~al.}(2016){Eistrup}, {Walsh}, \& {van
  Dishoeck}}]{Eistrup2016}
{Eistrup}, C., {Walsh}, C., \& {van Dishoeck}, E.~F. 2016, \aap, 595, A83

\bibitem[{{Eistrup} {et~al.}(2018){Eistrup}, {Walsh}, \& {van
  Dishoeck}}]{Eistrup2018}
{Eistrup}, C., {Walsh}, C., \& {van Dishoeck}, E.~F. 2018, \aap, 613, A14

\bibitem[{{Fung} {et~al.}(2019){Fung}, {Zhu}, \& {Chiang}}]{Fung2019}
{Fung}, J., {Zhu}, Z., \& {Chiang}, E. 2019, \apj, 887, 152

\bibitem[{{Gibb} \& {Horne}(2013)}]{Gibb2013}
{Gibb}, E.~L. \& {Horne}, D. 2013, \apjl, 776, L28

\bibitem[{{G{\"u}ver} \& {{\"O}zel}(2009)}]{TolgaOzel2009}
{G{\"u}ver}, T. \& {{\"O}zel}, F. 2009, \mnras, 400, 2050

\bibitem[{{Haffert} {et~al.}(2019){Haffert}, {Bohn}, {de Boer}, {Snellen},
  {Brinchmann}, {Girard}, {Keller}, \& {Bacon}}]{Haffert2019}
{Haffert}, S.~Y., {Bohn}, A.~J., {de Boer}, J., {et~al.} 2019, Nature
  Astronomy, 3, 749

\bibitem[{{Hartmann} {et~al.}(1998){Hartmann}, {Calvet}, {Gullbring}, \&
  {D'Alessio}}]{Hartmann1998}
{Hartmann}, L., {Calvet}, N., {Gullbring}, E., \& {D'Alessio}, P. 1998, \apj,
  495, 385

\bibitem[{Heller \& Pudritz(2015)}]{Heller_2015}
Heller, R. \& Pudritz, R. 2015, The Astrophysical Journal, 806, 181

\bibitem[{{Hunter}(2007)}]{matplotlib}
{Hunter}, J.~D. 2007, Computing in Science and Engineering, 9, 90

\bibitem[{{Jiang} {et~al.}(2023){Jiang}, {Wang}, {Ormel}, {Krijt}, \&
  {Dong}}]{Jiang2023}
{Jiang}, H., {Wang}, Y., {Ormel}, C.~W., {Krijt}, S., \& {Dong}, R. 2023, \aap,
  678, A33

\bibitem[{{Kaufman} \& {Neufeld}(1996)}]{Kaufman1996}
{Kaufman}, M.~J. \& {Neufeld}, D.~A. 1996, \apj, 456, 611

\bibitem[{{Keppler} {et~al.}(2018){Keppler}, {Benisty}, {M{\"u}ller},
  {Henning}, {van Boekel}, {Cantalloube}, {Ginski}, {van Holstein}, {Maire},
  {Pohl}, {Samland}, {Avenhaus}, {Baudino}, {Boccaletti}, {de Boer},
  {Bonnefoy}, {Chauvin}, {Desidera}, {Langlois}, {Lazzoni}, {Marleau},
  {Mordasini}, {Pawellek}, {Stolker}, {Vigan}, {Zurlo}, {Birnstiel},
  {Brandner}, {Feldt}, {Flock}, {Girard}, {Gratton}, {Hagelberg}, {Isella},
  {Janson}, {Juhasz}, {Kemmer}, {Kral}, {Lagrange}, {Launhardt}, {Matter},
  {M{\'e}nard}, {Milli}, {Molli{\`e}re}, {Olofsson}, {P{\'e}rez}, {Pinilla},
  {Pinte}, {Quanz}, {Schmidt}, {Udry}, {Wahhaj}, {Williams}, {Buenzli},
  {Cudel}, {Dominik}, {Galicher}, {Kasper}, {Lannier}, {Mesa}, {Mouillet},
  {Peretti}, {Perrot}, {Salter}, {Sissa}, {Wildi}, {Abe}, {Antichi},
  {Augereau}, {Baruffolo}, {Baudoz}, {Bazzon}, {Beuzit}, {Blanchard}, {Brems},
  {Buey}, {De Caprio}, {Carbillet}, {Carle}, {Cascone}, {Cheetham}, {Claudi},
  {Costille}, {Delboulb{\'e}}, {Dohlen}, {Fantinel}, {Feautrier}, {Fusco},
  {Giro}, {Gluck}, {Gry}, {Hubin}, {Hugot}, {Jaquet}, {Le Mignant}, {Llored},
  {Madec}, {Magnard}, {Martinez}, {Maurel}, {Meyer}, {M{\"o}ller-Nilsson},
  {Moulin}, {Mugnier}, {Orign{\'e}}, {Pavlov}, {Perret}, {Petit}, {Pragt},
  {Puget}, {Rabou}, {Ramos}, {Rigal}, {Rochat}, {Roelfsema}, {Rousset}, {Roux},
  {Salasnich}, {Sauvage}, {Sevin}, {Soenke}, {Stadler}, {Suarez}, {Turatto}, \&
  {Weber}}]{Keppler2018}
{Keppler}, M., {Benisty}, M., {M{\"u}ller}, A., {et~al.} 2018, \aap, 617, A44

\bibitem[{{Keyte} {et~al.}(2023){Keyte}, {Kama}, {Booth}, {Bergin}, {Cleeves},
  {van Dishoeck}, {Drozdovskaya}, {Furuya}, {Rawlings}, {Shorttle}, \&
  {Walsh}}]{Keyte2023}
{Keyte}, L., {Kama}, M., {Booth}, A.~S., {et~al.} 2023, Nature Astronomy, 7,
  684

\bibitem[{{Klahr} \& {Kley}(2006)}]{KlahrKley2006}
{Klahr}, H. \& {Kley}, W. 2006, \aap, 445, 747

\bibitem[{{Krapp} {et~al.}(2024){Krapp}, {Kratter}, {Youdin},
  {Ben{\'\i}tez-Llambay}, {Masset}, \& {Armitage}}]{2024arXiv240214638K}
{Krapp}, L., {Kratter}, K.~M., {Youdin}, A.~N., {et~al.} 2024, arXiv e-prints,
  arXiv:2402.14638

\bibitem[{{Krijt} {et~al.}(2020){Krijt}, {Bosman}, {Zhang}, {Schwarz},
  {Ciesla}, \& {Bergin}}]{Krijt2020}
{Krijt}, S., {Bosman}, A.~D., {Zhang}, K., {et~al.} 2020, \apj, 899, 134

\bibitem[{{Lambrechts} \& {Lega}(2017)}]{LL17}
{Lambrechts}, M. \& {Lega}, E. 2017, \aap, 606, A146

\bibitem[{{Lambrechts} {et~al.}(2019){Lambrechts}, {Lega}, {Nelson}, {Crida},
  \& {Morbidelli}}]{LLNCM19}
{Lambrechts}, M., {Lega}, E., {Nelson}, R.~P., {Crida}, A., \& {Morbidelli}, A.
  2019, \aap, 630, A82

\bibitem[{{Law} {et~al.}(2023){Law}, {Booth}, \& {{\"O}berg}}]{Law2023}
{Law}, C.~J., {Booth}, A.~S., \& {{\"O}berg}, K.~I. 2023, \apjl, 952, L19

\bibitem[{{Lega} {et~al.}(2024){Lega}, {Benisty}, {Cridland}, {Morbidelli},
  {Schulik}, \& {Lambrechts}}]{Legaetal}
{Lega}, E., {Benisty}, M., {Cridland}, A., {et~al.} 2024, arXiv e-prints,
  arXiv:2408.12233

\bibitem[{{Lega} {et~al.}(2014){Lega}, {Crida}, {Bitsch}, \&
  {Morbidelli}}]{2014MNRAS.440..683L}
{Lega}, E., {Crida}, A., {Bitsch}, B., \& {Morbidelli}, A. 2014, \mnras, 440,
  683

\bibitem[{{Marleau} {et~al.}(2023){Marleau}, {Kuiper}, {B{\'e}thune}, \&
  {Mordasini}}]{Marleau2023}
{Marleau}, G.-D., {Kuiper}, R., {B{\'e}thune}, W., \& {Mordasini}, C. 2023,
  \apj, 952, 89

\bibitem[{{Mathis} {et~al.}(1977){Mathis}, {Rumpl}, \& {Nordsieck}}]{MRN77}
{Mathis}, J.~S., {Rumpl}, W., \& {Nordsieck}, K.~H. 1977, \apj, 217, 425

\bibitem[{{McClure} {et~al.}(2023){McClure}, {Rocha}, {Pontoppidan}, {Crouzet},
  {Chu}, {Dartois}, {Lamberts}, {Noble}, {Pendleton}, {Perotti}, {Qasim},
  {Rachid}, {Smith}, {Sun}, {Beck}, {Boogert}, {Brown}, {Caselli}, {Charnley},
  {Cuppen}, {Dickinson}, {Drozdovskaya}, {Egami}, {Erkal}, {Fraser}, {Garrod},
  {Harsono}, {Ioppolo}, {Jim{\'e}nez-Serra}, {Jin}, {J{\o}rgensen},
  {Kristensen}, {Lis}, {McCoustra}, {McGuire}, {Melnick}, {{\~A}-berg},
  {Palumbo}, {Shimonishi}, {Sturm}, {van Dishoeck}, \&
  {Linnartz}}]{McClure2023}
{McClure}, M.~K., {Rocha}, W.~R.~M., {Pontoppidan}, K.~M., {et~al.} 2023,
  Nature Astronomy, 7, 431

\bibitem[{{Meijerink} {et~al.}(2009){Meijerink}, {Pontoppidan}, {Blake},
  {Poelman}, \& {Dullemond}}]{Meijerink2009}
{Meijerink}, R., {Pontoppidan}, K.~M., {Blake}, G.~A., {Poelman}, D.~R., \&
  {Dullemond}, C.~P. 2009, \apj, 704, 1471

\bibitem[{{Miotello} {et~al.}(2019){Miotello}, {Facchini}, {van Dishoeck},
  {Cazzoletti}, {Testi}, {Williams}, {Ansdell}, {van Terwisga}, \& {van der
  Marel}}]{Miotello2019}
{Miotello}, A., {Facchini}, S., {van Dishoeck}, E.~F., {et~al.} 2019, \aap,
  631, A69

\bibitem[{{Morbidelli} {et~al.}(2014){Morbidelli}, {Szul{\'a}gyi}, {Crida},
  {Lega}, {Bitsch}, {Tanigawa}, \& {Kanagawa}}]{Morbidelli2014}
{Morbidelli}, A., {Szul{\'a}gyi}, J., {Crida}, A., {et~al.} 2014, \icarus, 232,
  266

\bibitem[{{Mordasini}(2013)}]{Mordasini2013}
{Mordasini}, C. 2013, \aap, 558, A113

\bibitem[{{Rampinelli} {et~al.}(2024){Rampinelli}, {Facchini}, {Leemker},
  {Bae}, {Benisty}, {Teague}, {Law}, {{\"O}berg}, {Portilla-Revelo}, \&
  {Cridland}}]{Rampinelli2024}
{Rampinelli}, L., {Facchini}, S., {Leemker}, M., {et~al.} 2024, \aap, 689, A65

\bibitem[{{Semenov} {et~al.}(2018){Semenov}, {Favre}, {Fedele}, {Guilloteau},
  {Teague}, {Henning}, {Dutrey}, {Chapillon}, {Hersant}, \&
  {Pi{\'e}tu}}]{Semenov2018}
{Semenov}, D., {Favre}, C., {Fedele}, D., {et~al.} 2018, \aap, 617, A28

\bibitem[{{Semenov} {et~al.}(2010){Semenov}, {Hersant}, {Wakelam}, {Dutrey},
  {Chapillon}, {Guilloteau}, {Henning}, {Launhardt}, {Pi{\'e}tu}, \&
  {Schreyer}}]{Sem10}
{Semenov}, D., {Hersant}, F., {Wakelam}, V., {et~al.} 2010, \aap, 522, A42

\bibitem[{{Semenov}(2017)}]{Sem17}
{Semenov}, D.~A. 2017, {ALCHEMIC: Advanced time-dependent chemical kinetics},
  Astrophysics Source Code Library, record ascl:1708.008

\bibitem[{{Shibaike} \& {Mori}(2023)}]{Shibaike2023}
{Shibaike}, Y. \& {Mori}, S. 2023, \mnras, 518, 5444

\bibitem[{{Stolker} {et~al.}(2020){Stolker}, {Marleau}, {Cugno},
  {Molli{\`e}re}, {Quanz}, {Todorov}, \& {K{\"u}hn}}]{Stolker2020}
{Stolker}, T., {Marleau}, G.~D., {Cugno}, G., {et~al.} 2020, \aap, 644, A13

\bibitem[{{Szul{\'a}gyi} {et~al.}(2016){Szul{\'a}gyi}, {Masset}, {Lega},
  {Crida}, {Morbidelli}, \& {Guillot}}]{Szulagyi2016}
{Szul{\'a}gyi}, J., {Masset}, F., {Lega}, E., {et~al.} 2016, \mnras, 460, 2853

\bibitem[{{Szul{\'a}gyi} {et~al.}(2014){Szul{\'a}gyi}, {Morbidelli}, {Crida},
  \& {Masset}}]{Szul2014}
{Szul{\'a}gyi}, J., {Morbidelli}, A., {Crida}, A., \& {Masset}, F. 2014, \apj,
  782, 65

\bibitem[{{Tabone} {et~al.}(2023){Tabone}, {Bettoni}, {van Dishoeck},
  {Arabhavi}, {Grant}, {Gasman}, {Henning}, {Kamp}, {G{\"u}del}, {Lagage},
  {Ray}, {Vandenbussche}, {Abergel}, {Absil}, {Argyriou}, {Barrado},
  {Boccaletti}, {Bouwman}, {Caratti o Garatti}, {Geers}, {Glauser},
  {Justannont}, {Lahuis}, {Mueller}, {Nehm{\'e}}, {Olofsson}, {Pantin},
  {Scheithauer}, {Waelkens}, {Waters}, {Black}, {Christiaens}, {Guadarrama},
  {Morales-Calder{\'o}n}, {Jang}, {Kanwar}, {Pawellek}, {Perotti}, {Perrin},
  {Rodgers-Lee}, {Samland}, {Schreiber}, {Schwarz}, {Colina}, {{\"O}stlin}, \&
  {Wright}}]{Tabone2023}
{Tabone}, B., {Bettoni}, G., {van Dishoeck}, E.~F., {et~al.} 2023, Nature
  Astronomy, 7, 805

\bibitem[{{van der Walt} {et~al.}(2011){van der Walt}, {Colbert}, \&
  {Varoquaux}}]{numpy}
{van der Walt}, S., {Colbert}, S.~C., \& {Varoquaux}, G. 2011, Computing in
  Science and Engineering, 13, 22

\bibitem[{{van Dishoeck} \& {Black}(1982)}]{vDB1982}
{van Dishoeck}, E.~F. \& {Black}, J.~H. 1982, \apj, 258, 533

\bibitem[{{van Dishoeck} {et~al.}(2021{\natexlab{a}}){van Dishoeck},
  {Kristensen}, {Mottram}, {Benz}, {Bergin}, {Caselli}, {Herpin},
  {Hogerheijde}, {Johnstone}, {Liseau}, {Nisini}, {Tafalla}, {van der Tak},
  {Wyrowski}, {Baudry}, {Benedettini}, {Bjerkeli}, {Blake}, {Braine},
  {Bruderer}, {Cabrit}, {Cernicharo}, {Choi}, {Coutens}, {de Graauw},
  {Dominik}, {Fedele}, {Fich}, {Fuente}, {Furuya}, {Goicoechea}, {Harsono},
  {Helmich}, {Herczeg}, {Jacq}, {Karska}, {Kaufman}, {Keto}, {Lamberts},
  {Larsson}, {Leurini}, {Lis}, {Melnick}, {Neufeld}, {Pagani}, {Persson},
  {Shipman}, {Taquet}, {van Kempen}, {Walsh}, {Wampfler}, {Y{\i}ld{\i}z}, \&
  {WISH Team}}]{vanDishoeck2021}
{van Dishoeck}, E.~F., {Kristensen}, L.~E., {Mottram}, J.~C., {et~al.}
  2021{\natexlab{a}}, \aap, 648, A24

\bibitem[{{van Dishoeck} {et~al.}(2021{\natexlab{b}}){van Dishoeck},
  {Kristensen}, {Mottram}, {Benz}, {Bergin}, {Caselli}, {Herpin},
  {Hogerheijde}, {Johnstone}, {Liseau}, {Nisini}, {Tafalla}, {van der Tak},
  {Wyrowski}, {Baudry}, {Benedettini}, {Bjerkeli}, {Blake}, {Braine},
  {Bruderer}, {Cabrit}, {Cernicharo}, {Choi}, {Coutens}, {de Graauw},
  {Dominik}, {Fedele}, {Fich}, {Fuente}, {Furuya}, {Goicoechea}, {Harsono},
  {Helmich}, {Herczeg}, {Jacq}, {Karska}, {Kaufman}, {Keto}, {Lamberts},
  {Larsson}, {Leurini}, {Lis}, {Melnick}, {Neufeld}, {Pagani}, {Persson},
  {Shipman}, {Taquet}, {van Kempen}, {Walsh}, {Wampfler}, {Y{\i}ld{\i}z}, \&
  {WISH Team}}]{WISHreview2021}
{van Dishoeck}, E.~F., {Kristensen}, L.~E., {Mottram}, J.~C., {et~al.}
  2021{\natexlab{b}}, \aap, 648, A24

\bibitem[{{van Gelder} {et~al.}(2021){van Gelder}, {Tabone}, {van Dishoeck}, \&
  {Godard}}]{vanGelder2021}
{van Gelder}, M.~L., {Tabone}, B., {van Dishoeck}, E.~F., \& {Godard}, B. 2021,
  \aap, 653, A159

\bibitem[{{Virtanen} {et~al.}(2020){Virtanen}, {Gommers}, {Burovski},
  {Oliphant}, {Weckesser}, {Cournapeau}, {alexbrc}, {Peterson}, {Reddy},
  {Wilson}, {Haberland}, {Mayorov}, {endolith}, {Nelson}, {van der Walt},
  {Laxalde}, {Brett}, {Polat}, {Larson}, {Millman}, {Lars}, {van Mulbregt},
  {eric-jones}, {Carey}, {Moore}, {Kern}, {Leslie}, {Perktold}, {Striega}, \&
  {Feng}}]{Scipy}
{Virtanen}, P., {Gommers}, R., {Burovski}, E., {et~al.} 2020, {scipy/scipy:
  SciPy 1.5.3}, Zenodo

\bibitem[{{Visser} {et~al.}(2018){Visser}, {Bruderer}, {Cazzoletti},
  {Facchini}, {Heays}, \& {van Dishoeck}}]{Visser2018}
{Visser}, R., {Bruderer}, S., {Cazzoletti}, P., {et~al.} 2018, \aap, 615, A75

\bibitem[{{Wakelam} {et~al.}(2012){Wakelam}, {Herbst}, {Loison}, {Smith},
  {Chandrasekaran}, {Pavone}, {Adams}, {Bacchus-Montabonel}, {Bergeat},
  {B{\'e}roff}, {Bierbaum}, {Chabot}, {Dalgarno}, {van Dishoeck}, {Faure},
  {Geppert}, {Gerlich}, {Galli}, {H{\'e}brard}, {Hersant}, {Hickson},
  {Honvault}, {Klippenstein}, {Le Picard}, {Nyman}, {Pernot}, {Schlemmer},
  {Selsis}, {Sims}, {Talbi}, {Tennyson}, {Troe}, {Wester}, \&
  {Wiesenfeld}}]{Wakelam2012}
{Wakelam}, V., {Herbst}, E., {Loison}, J.~C., {et~al.} 2012, \apjs, 199, 21

\bibitem[{{Walsh} {et~al.}(2015){Walsh}, {Nomura}, \& {van Dishoeck}}]{Walsh15}
{Walsh}, C., {Nomura}, H., \& {van Dishoeck}, E. 2015, \aap, 582, A88

\bibitem[{{Wang} {et~al.}(2020){Wang}, {Leigh}, {Perna}, \& {Shara}}]{Wang2020}
{Wang}, Y.-H., {Leigh}, N. W.~C., {Perna}, R., \& {Shara}, M.~M. 2020, \apj,
  905, 136

\bibitem[{{Wolff} {et~al.}(2017){Wolff}, {M{\'e}nard}, {Caceres},
  {Lef{\`e}vre}, {Bonnefoy}, {C{\'a}novas}, {Maret}, {Pinte}, {Schreiber}, \&
  {van der Plas}}]{Wolff2017}
{Wolff}, S.~G., {M{\'e}nard}, F., {Caceres}, C., {et~al.} 2017, \aj, 154, 26

\bibitem[{{Zuckerman} {et~al.}(1976){Zuckerman}, {Kuiper}, \& {Rodriguez
  Kuiper}}]{Zuckerman1976}
{Zuckerman}, B., {Kuiper}, T.~B.~H., \& {Rodriguez Kuiper}, E.~N. 1976, \apjl,
  209, L137

\end{thebibliography}

\appendix

\section{Supplementary figures} \label{sec:appendix}

\subsection{Streamline density evolution}

\begin{figure}
    \centering
    \includegraphics[width=0.8\linewidth]{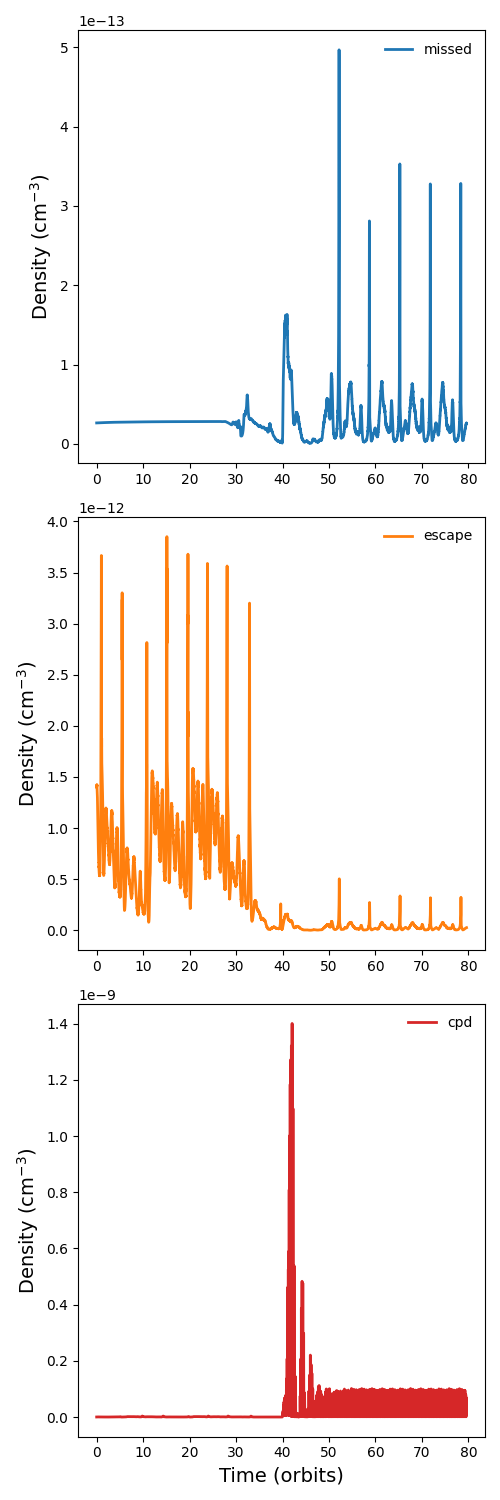}
    \caption{The density evolution for the three test streamlines in the LowMass model. The disk streamline (red) experiences an increase in gas density by about 3 orders of magntiude when it enters into the CPD.}
    \label{fig:dens}
\end{figure}

In figure \ref{fig:dens} we show the density evolution of streamlines mainly in the PPD (missed and escape; blue and orange, respectively) and the one that enters into the CPD (disk; red). There is a significant difference in the gas volume density for the disk streamline which alter the steady state sublimation temperature of volatiles by increasing the rate at which the gas can re-collide with the dust grains.

\subsection{Chemical evolution zoom-in}

\begin{figure*}
    \centering
    \includegraphics[width=0.8\textwidth]{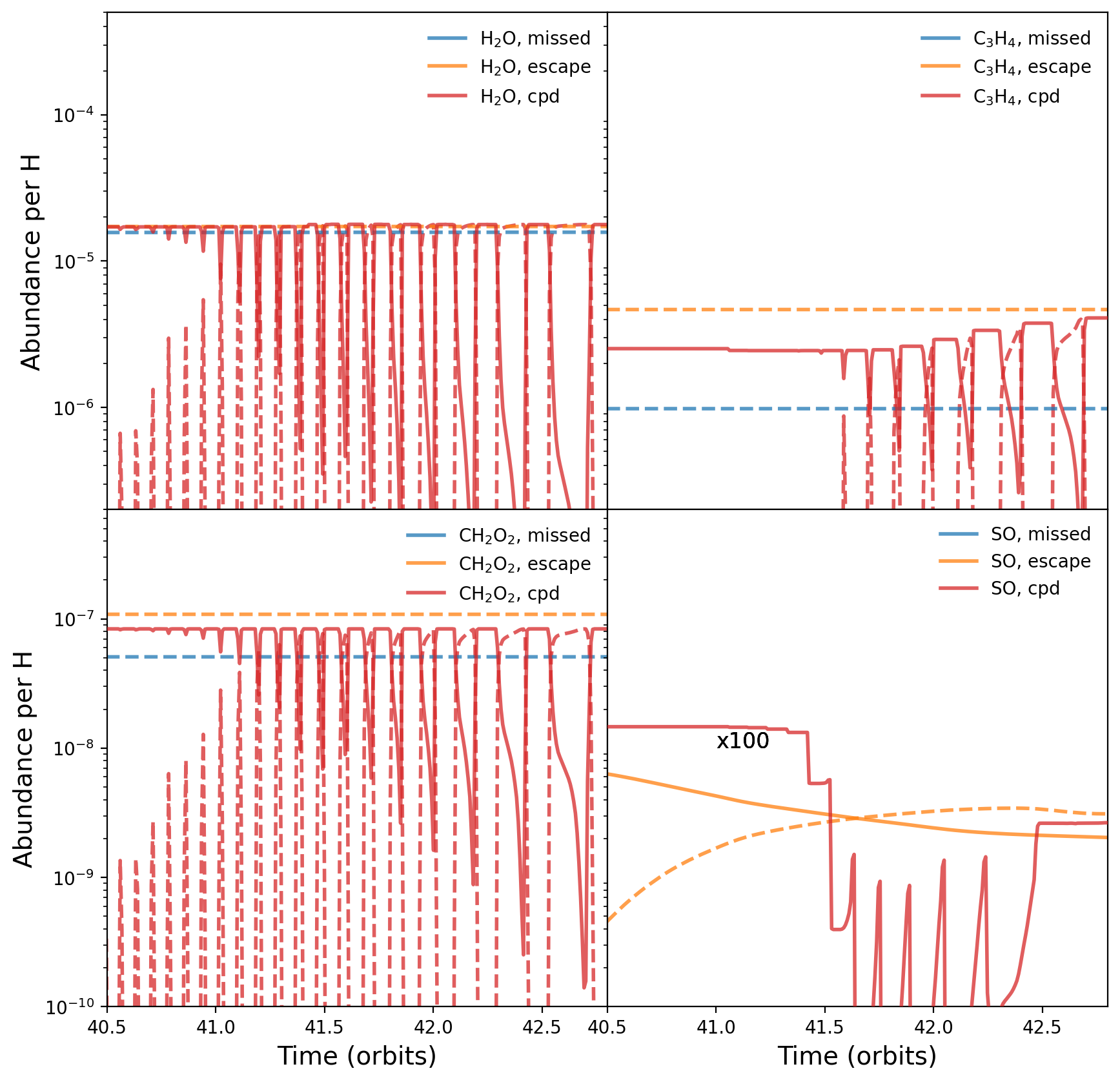}
    \caption{A zoom in view of the chemical evolution of each streamline just after $t=40$ orbits, or when the streamlines undergo their closest approach to the embedded planet, and disk streamline enters the CPD. The selected species are the ones that undergo oscillation in the CPD as the streamline orbits in and out of warm CPD regions.} 
    \label{fig:zoom}
\end{figure*}

In figure \ref{fig:zoom} we show a zoomed-in view of the chemistry of the streamlines near where they had their closest approach to the embedded planet, and when the disk streamline (red) enters the CPD. We focus mainly on the first few orbits after the disk streamline enters the CPD where the density (figure \ref{fig:dens} and temperatures (figure \ref{fig:temp}) are the highest.

There we still see an oscillation between gas (solid) and ice (dashed) phases for these species and as the disk streamline cycles away from the embedded planet, to cooler temperatures the fraction of ice/gas approaches a state where the two phases oscillate perfectly between one and the other. Species like SO, which require OH in the gas phase to be formed also oscillate as the source of OH, H$_2$O vapour oscillates between its gas and ice phase.
\end{document}